\documentclass{aa}  

\usepackage{xcolor}
\usepackage{bm}

\usepackage{graphicx}
\usepackage{txfonts}
\newcommand{\Mpc}{\, \rm{Mpc} }

\newcommand{\Msun}{\, \rm{M_{\odot}} }

\begin{document}

   \title{Comparison of Bayesian inference methods using the Loreli II database of hydro-radiative simulations of the 21-cm signal}

   \author{R.Meriot
          \inst{1},
          B.Semelin\inst{1},
          \and D.Cornu\inst{1}
          }

   \institute{ Observatoire de Paris, PSL Research University,  Sorbonne Université, CNRS, LERMA, 75014 Paris, France}

   \date{}

 
  \abstract
   {While the observation of the 21 cm signal from the Cosmic Dawn and Epoch of Reionization is an instrumental challenge, the interpretation of a prospective detection is still open to questions regarding the modelling of the signal and the bayesian inference techniques that bridge the gap between theory and observations. To address some of these questions, we present Loreli II, a database of nearly 10 000 simulations of the 21 cm signal run with the Licorice 3D radiative transfer code. With Loreli II, we explore a 5-dimensional astrophysical parameter space where star formation, X-ray emissions, and UV emissions are varied. We then use this database to train neural networks and perform bayesian inference on 21 cm power spectra affected by thermal noise at the level of 100 hours of observation with the  Square Kilometer Array. We study and compare three inference techniques : an emulator of the power spectrum, a Neural Density Estimator that fits the implicit likelihood of the model, and a Bayesian Neural Network that directly fits the posterior distribution. We measure the performances of each method by comparing them on a statistically representative set of inferences, notably using the principles of Simulation-Based Calibration. We report errors on the 1-D marginalized posteriors (biases and over/under confidence) below $15 \%$ of the standard deviation for the emulator and below $25 \%$ for the other methods. We conclude that at our noise level and our sampling density of the parameter space, an explicit Gaussian likelihood is sufficient. This may not be the case at lower noise level or if a denser sampling is used to reach higher accuracy. We then apply the emulator method to recent HERA upper limits and report weak constraints on the X-ray emissivity parameter of our model. }


   \keywords{Cosmology: dark ages, reionization, first stars --  Radiative transfer-- Early Universe -- Methods: numerical
                              }

\titlerunning{Bayesian inference with Loreli II simulations}

   \maketitle
   
%
\section{Introduction}

The 21-cm signal of neutral hydrogen is a promising probe of the early universe. Emitted by the intergalactic medium (IGM), its features are shaped by the properties of the first structures during the first billion years of the Universe. Indeed, as the first galaxies appear during the so-called Cosmic Dawn ($z \gtrapprox 20 $), they heat and ionize their surrounding. Due to the Wouthuysen-Field effect \citep{Wouthuysen1952, Field1958}, the spin temperature of neutral Hydrogen leaves a state of equilibrium with the Cosmic Microwave Background (CMB). Through the action of the local Lyman-$\alpha$ flux, it becomes coupled with the local kinetic temperature. As a consequence, a contrast between the brightness temperature of the CMB and the 21-cm signal appears. Its intensity depends mainly on the local density of neutral Hydrogen and on its temperature, and is therefore linked to the UV and X-ray emissions of the first galaxies. Measuring the 21-cm signal would therefore allow astrophysicists to put tight constraints over the physical properties of the first sources of light such as the minimum mass of star forming dark matter halos or the spectral emissivity of the galaxies in the UV and X-ray bands.

Accessing this wealth of information remains an observational challenge. The Square Kilometer Array should be able to accurately measure the signal power spectrum and eventually provide a full 3D tomography of the 21-cm signal. In the meantime, current radio-interferometers (e.g. LOFAR, NenuFAR, HERA, MWA) are attempting to constrain the power spectrum of the signal, and are reporting ever-tightening upper limits on this statistical quantity \citep[e.g.][]{Abdurashidova2022a, Mertens2020, Munshi2023, Yoshiura2021}. However, the 21-cm signal is estimated to be around $10^4$ times fainter than the foreground emission of nearby radio sources. These must therefore be modelled with extreme care before being subtracted from the observations, or altogether avoided,  sacrificing part of the information in the process. Systematics of the instrument must also be mastered to bring the detection level to its theoretical limit, the thermal noise.  
In parallel, single-dipole experiments try to detect the average value of the 21-cm signal in the whole sky. EDGES reported a very deep signal \citep[$\sim -500$ mK at $z\sim 20$]{Bowmana} , a result that subsequent experiments failed to reproduce \citep{Singh2021} and that is not in the range of signals predicted with standard models of reionization. 

Producing accurate models of the 21-cm signal during the first billion years of the Universe is equally challenging. The physical processes at play couple scales across many orders of magnitude, forcing numerical setups to simulate $\gtrapprox 300 \Mpc$ boxes while taking into account galactic-scale phenomena. One also has to account for the propagation of UV and X-ray photons and their interaction with the IGM. To deal with this, the community chose a variety of approaches. The first is to use time-saving approximations whenever possible, resulting in a number of semi-analytical codes \citep[see e.g.][]{Sobacchi2014, Reis2021}. Those codes typically simulate the evolution of the dark matter density field, statistically estimate the population of dark matter halos in each resolution element and, crucially, bypass the actual simulation of the radiative transfer by estimating the size of ionized regions surrounding sources using photon budget arguments. This allows a computation of a single 21-cm field in a couple CPU hours (cpuh) in typical setups. On the opposite side of the spectrum we find fully numerical hydro-radiative simulations that are, comparatively, extremely computationally expensive, as these codes can require millions of cpuh to simulate a single box at high resolution. They are assumed to be more accurate, as they tend to rely on fewer approximations and more on first principle physics. They typically differentiate between the dynamics of dark matter and baryonic matter, resolve the relevant gravitationally bound structures, and perform 3D radiative transfer. Among these codes is Licorice, which is used in this work as well as in e.g. \cite{Meriot2024} (see also references therein).  The community also designed a few codes of intermediate cost, such as \cite{Schaeffer2023,  Ghara2018}, which typically "paint" baryons over simulated dark matter density fields or halo catalogues, and solve spherically symmetric 1D radiative transfer to compute the 21-cm signal. 

The interpretation of instrumental data using Bayesian Inference typically requires tens or hundreds of thousands of realizations with different sets of astrophysical parameters and, ideally, many realizations of the density field at each point in parameter space. This makes the method computationally intractable for codes that take more than a few cpuh to simulate the EoR. For this reason, semi-analytical codes have been widely used to develop data processing and inference pipelines. However, in a previous work \citep{Meriot2024}, we presented Loreli I, a database of 760 moderate-resolution 3D radiative transfer simulations of the EoR suited for inference purposes.

The most common inference technique used by the community is the classical Monte Carlo Markov Chain approach. While its convergence toward the true posterior distribution is guaranteed, it suffers from two caveats. First, it requires an explicitly written likelihood, which in general does not exist for 21-cm summary statistics and exists only in an approximate form for the 21-cm power spectrum. Second, it requires the simulations to be run sequentially during the inference process and not upfront: performing inference on $n$ different observed data sets requires $n$ times the computation cost needed for one data set, and the process is said to be "non-amortized". In practice, it is only suited for the fastest codes. This led the community to explore alternative methods based on machine learning. For instance, one consists in performing Simulation Based Inference (SBI). As opposed to the MCMC algorithm, SBI does not require an explicit likelihood: instead, it takes advantage of the fact that a result from the modelling pipeline (which typically includes the generation of the initial density field, simulation of the cosmological 21-cm signal and modelling of the instrumental noise) is a draw in the likelihood inherent to the model. Given that the implicit likelihood defined by the model is not subject to approximations, that the only requirement for SBI is to be able to generate signals and that the computation cost is upfront and amortized, this approach has received significant attention in the 21-cm community \citep[see e.g.][]{Papamakarios, Alsing2018, Alsing2019, Zhao2022, Prelogovic2023}. Different versions of SBI exists, here we focus on training a Neural Density Estimator (NDE) to fit the likelihood before using the fit in an MCMC pipeline, and training a Bayesian Neural Network (BNN) to estimate the posterior.

In this paper, we greatly expand on the work presented in \cite{Meriot2024}. In section 1,  we present Loreli II, a database of 10 000 hydro-radiative numerical simulations of the EoR. In section 2, we give details about the three machine-learning based inference techniques that we explore in this work, namely using of an emulator of the simulation code in an MCMC pipeline, using a Neural Density Estimator to fit the true likelihood sampled by the training set, and using a Bayesian Neural Network to directly predict the posterior. In section 3, we present the performance of these inference methods and comment on their convergence.
\section{The LoReLi II database}

\subsection{The Licorice simulation code}

Here we briefly summarize the features of the simulation code, and present the new version of the Loreli database. Licorice is an N-body hydro-radiative simulation code of the EoR. Dynamics of gas and dark matter particles is solved using a TREE+SPH approach \citep{Appel1985, Barnes1986, Springel2005}. As detailed in \cite{Meriot2024}, the conditional mass function (CMF) of each gas particle is calculated at every time step, following the extension of the Sheth-Tormen mass function \citep{Sheth} presented  in \cite{Rubino-Martin2008}. The CMF formalism gives a statistical estimate of the average number of unresolved dark matter halos in a given mass bin, in the region represented by a gas particle. After integrating the CMF between a minimum halo mass $M_{min}$ and the mass of the region, it is straightforward to obtain $f_{coll}$, the fraction of the mass of the region in collapsed structures i.e. dark matter halos. This quantity is directly used to compute $df_{*}$, the fraction of stellar mass created in each particle at each dynamical time step, according to 

 \begin{equation}\label{eq:sf}
     df_{*} = (f_{coll} - f_{*})\dfrac{dt}{\tau_{SF}},
 \end{equation}

 \noindent where $dt$ is the duration of a dynamical time step, $\tau$ a parameter of the model that represents the typical timescale over which collapsed gas is converted into stars, and $f_{*}$ the stellar fraction of a particle. Particles with non-zero $f_{*}$ are considered sources and emit photon packets in the UV and X-ray continuum.  To solve radiative transfer, gas particles are dispatched on an adaptive grid. Photon packets propagate at the speed of light and deposit energy in the grid cells they encounter. The temperature and ionization state of gas particles in these cells are then updated over radiative time steps that may be up to 10000 smaller than the dynamical time steps. A two phase model of the particles has been adopted: both the average temperature and the temperature of the neutral phase of the particles are computed, as calculating the 21 cm signal of a region using its average temperature leads to an overestimate  \citep[see][]{Meriot2024}. UV photon packets propagate until they are completely absorbed, but in order to save computing power, hard X-rays (with an energy greater than $2 \, \rm{keV}$, and a mean free path larger than the size of the box) are added to a uniform background after they propagated over distances larger than a box length. The simulation runs until the box is $99\%$ ionized. 

 The physical properties of the particles are then interpolated on a $256^3$ grid in a post-processing step. The Wouthuysen-Field coupling is computed on this grid using the semi-analytical code SPINTER \citep{Semelin2023}.  This allows for a computation of the brightness temperature of the 21 cm signal $\delta T_b$ using the classical equation (e.g., \cite{Furlanetto2006a}) at redshift $z$:

\begin{equation}\label{eq:dtb}
\begin{split}
        \delta T_b & = 27. \, x_{HI}(1+\delta) \left[ \frac{T_s - T_{CMB}}{T_s} \right] \left[ 1 + \frac{dv_{||}/dr_{||}}{ H(z) }  \right]^{-1}  \\ 
    & \times \left[ \frac{1+z}{10}\right]^{1/2} 
    \left[ \frac{\Omega_b}{0.044} \frac{h}{0.7} \right] 
    \left[ \frac{\Omega_m}{0.27}\right]^{1/2} \, \rm{mK}
 \end{split}
\end{equation}

\noindent where $x_{HI}$ is the local neutral Hydrogen fraction in a grid cell, $\delta$ the local overdensity, $H$ the Hubble parameter, $dv_{||}/dr_{||}$ the line-of-sight velocity gradient, $T_{CMB}$ the Cosmic Microwave Background temperature at redshift $z$, and $T_s$ the spin temperature of neutral Hydrogen. A complete description of the code can be found in \cite{Meriot2024} and references therein.

\subsection{The LoReLi II database}

Expanding upon LoReLi I, a database of 760 Licorice simulations presented in \cite{Meriot2024}, we now present LoReLi II, a set of 9828 simulations spanning a 5-dimensional parameter space. Simulations were run in $(300 \Mpc)^3 $ boxes with a $256^3$ resolution, with varying initial conditions. The cosmology was chosen to be consistent with Planck 2018 \citep{White2018}. The following parameters were varied: 

\begin{figure}
    \centering
    \includegraphics[scale = 0.3]{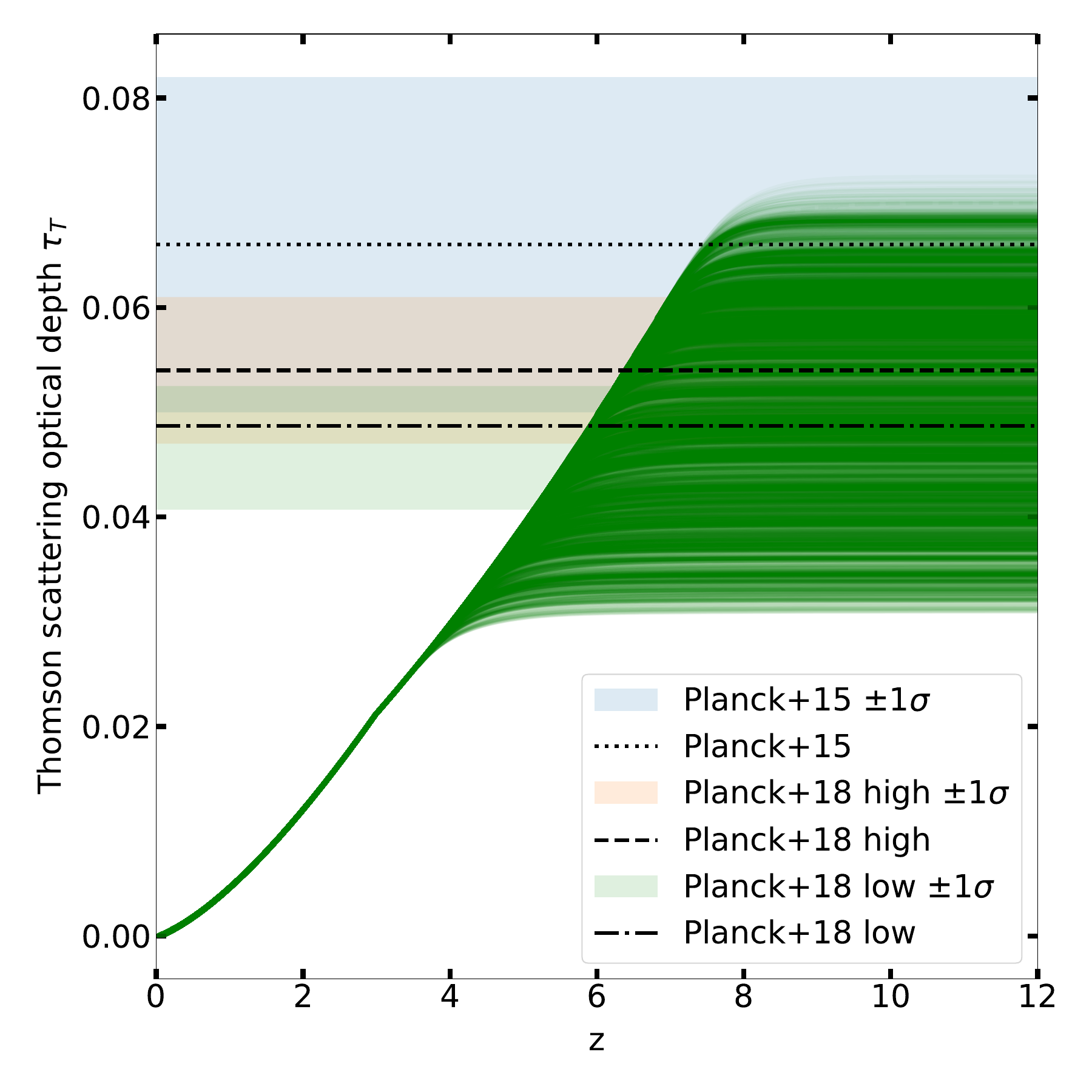}
    \caption{Thomson scattering optical depth of the \textsc{\textsc{Loreli}} II simulations. Measurements by Planck 2015 and 2018 are also reported \citep{White2018, Ade2016}. The \textsc{\textsc{Loreli}} simulations are within $3 \sigma$ of these constraints, although they were not used to calibrate the database per se.}
    \label{fig:optical_depth}
\end{figure}

\begin{itemize}
    \item The parameters of the subgrid star formation model : 42 pairs of $\tau_{SF} \in [7 \, Gyr , 105 \, Gyr  ] $ and $M_{min} \in [10^8 \, \Msun, 4 \times 10^9 \, \Msun]$ were selected (6 $\tau_{SF}$ values for each of the 7 $M_{min}$ values). They produce star formation rate densities within 2$\sigma$ of the observational data presented in \cite{Bouwens2016}, \cite{Oesch2018}, \cite{McLeod2016}.

    \item The escape fraction of UV radiation in particles with $\langle x_{HII} \rangle > 3\%$ : $f_{esc, post} \in \{0.05, 0.275, 0.5\}$. In particles with  $\langle x_{HII} \rangle < 3\%$, the escape fraction is kept at $0.003$.
    
    \item The X-ray production efficiency $f_X$ : 13 logarithmically spaced values in $[0.1,10]$. this parameter controls the X-ray luminosities of source particles according to  $L_X = 3.4 \times 10^{40} f_X \left(\frac{SFR}{1 \Msun yr^-1}\right) \rm{erg s^{-1} }$ (e.g. \cite{Furlanetto2006a})
    
    \item  The ratio between hard ($>2 \rm{keV}$) and soft ($<2 \rm{keV}$) X-ray $r_{H/S}$:  6 linearly spaced values in $[0,1]$. $r_{H/S}$ is the ratio of energy emitted by X-ray binaries to the total energy emitted in X-rays : $r_{H/S}   = \frac{f_X^{XRB}}{f_X}  $.
\end{itemize}

\begin{figure*}
        \includegraphics[scale = 0.35]{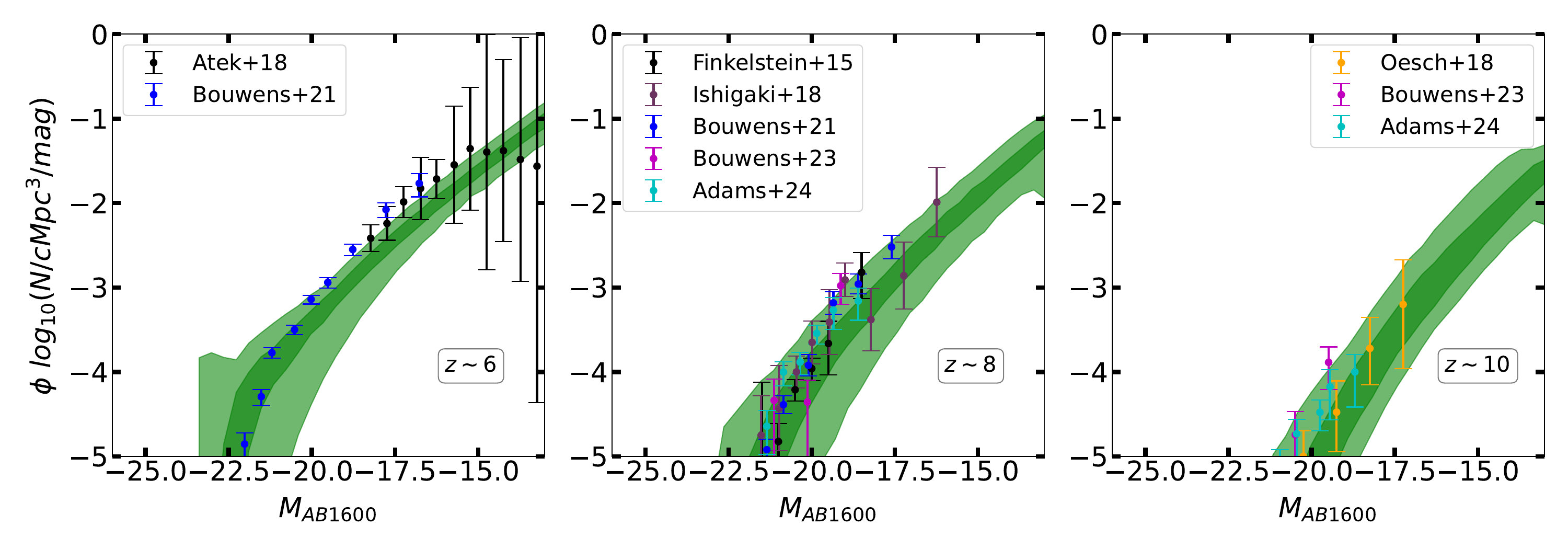}
  \caption{UV luminosity functions of the Loreli II simulations at $z\sim6, \, 8, \, 10$. The dark green region shows, for each magnitude bin, where the middle $50\%$ of the simulations lie, while the light green region shows where all simulations lie. We compare the simulated results with observations from \cite{Finkelstein2015}, \cite{Oesch2018}, \cite{Ishigaki2018}, \cite{Atek2018}, \cite{Bouwens2021}, \cite{Bouwens2023}. }
    \label{fig:uvlum10000}
\end{figure*}

In this setup, all simulations end between $z \lesssim 8 $ and $z \gtrsim 5 $, and all simulations have Thomson optical depths within $3 \sigma$ of Planck 2018 values, as shown in Fig. \ref{fig:optical_depth}. Similarly to Loreli I, Loreli II was initially calibrated on star formation observations. However, a combination of factors acting in opposite directions have skewed this calibration while mostly preserving the most important ingredient for the 21-cm signal: the cosmic emissivity of galaxies. Consequently, we present a comparison of the luminosity functions in Loreli to observed ones on Fig. \ref{fig:uvlum10000}. They were obtained in a post-processing step and we assumed the same SFR-to-halo-mass power-law index $\alpha = 5/3$ as was found in the CODA II simulation \citep[see][]{Ocvirk2020}. The simulated luminosity functions show an acceptable agreement with observed ones, although they on average lay slightly below $z\sim6$ constraints at $M_{AB1600}\gtrsim -20$, suggesting that the source model and star formation algorithm of Licorice may have to be updated in the future.

\begin{figure}
    \centering
    \includegraphics[scale = 0.25]{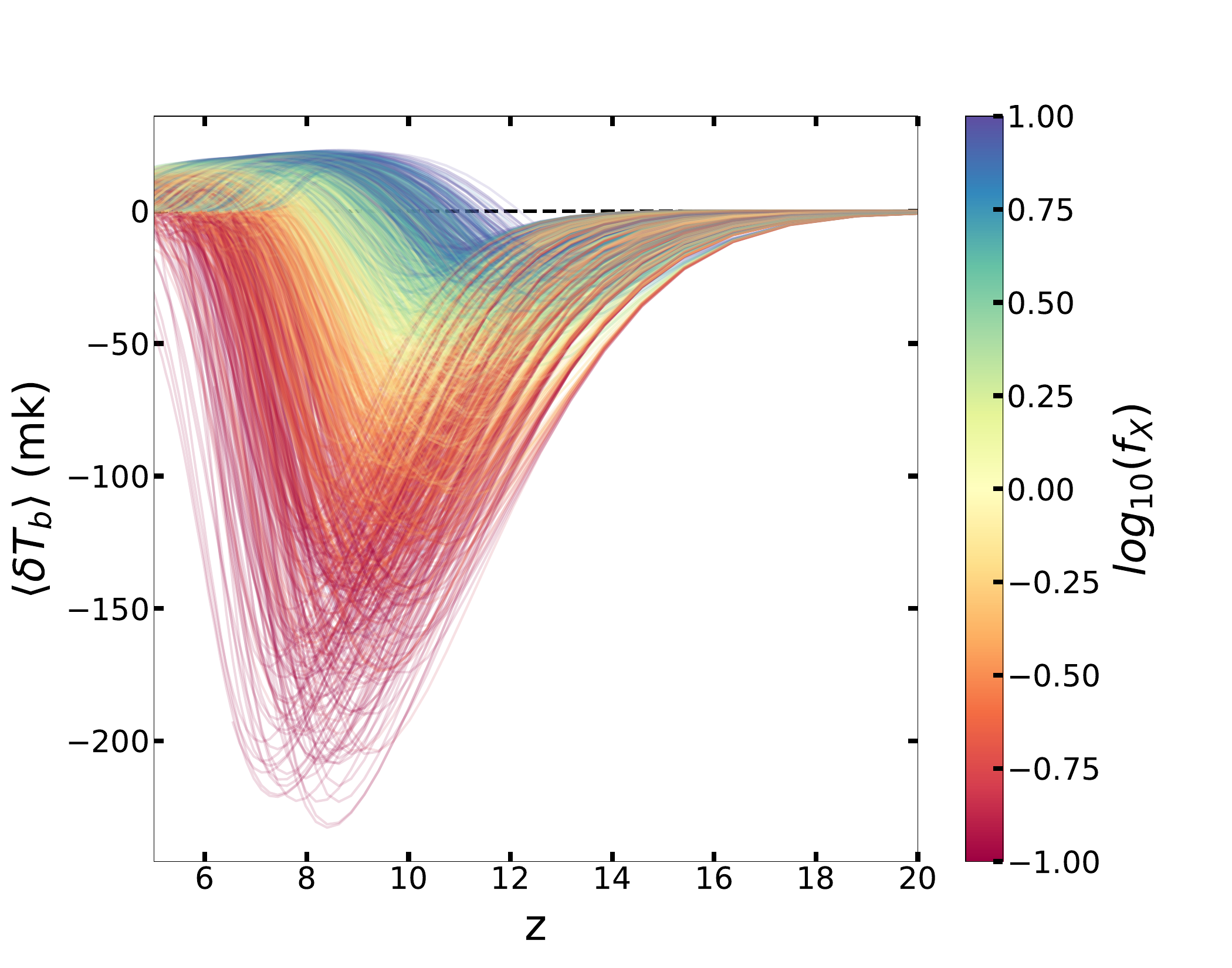}
    \caption{Average global $\delta T_b$ of the 10000 \textsc{\textsc{Loreli}} II simulations, colour coded according to the X-ray intensity parameter $f_X$. }
    \label{fig:dtb10000}
\end{figure}

The particle data have been saved at $\approx 50$ redshifts between $z \approx 60$ and the end of Reionization, resulting in approximately 1 PB of data. Another 0.5 PB consist in post-processed data, namely 3d grids of UV luminosities, Wouthuysen-Field coupling, ionization state,  21 cm brightness temperature, and its power spectrum. 
To illustrate the content of Loreli II, we show the global signals of Loreli II on Fig. \ref{fig:dtb10000}. The most intense signals reach $\sim -220 \, {\rm{mK}}$ at $z \sim 8$, far from the EDGES measurement at $-500 \, {\rm{mK}} $ at $z \sim 16$. The power spectra and global signals are publicly available at https://21ssd.obspm.fr/ .


\subsection{Sources of stochasticity}\label{sec:stoch}

Several stochastic processes affect the 21-cm power spectra. The first one is the  Cosmic Variance (CV, also called Sample Variance). Since a finite region of the sky is observed (or a finite volume is simulated), the larger the mode, the fewer times the power spectrum at that mode will be sampled. The measured value of the power spectrum will be an average of $N$ samplings of the underlying expected spectrum, and the variance on that average decreases as $N$ increases. While the small-scale modes will be extensively sampled and the resulting value will be very close to the theoretical average of the power spectrum, this introduces a significant uncertainty at large scales. Following \cite{Mcquinn} we can compute the covariance matrix of the CV as

\begin{equation}\label{eq:cv_noise}
     C_{CV}(k_i, k_j) \approx     P_{21}(k_i)\frac{\lambda^2 B^2}{Ax^2 y }\delta_{ij}
\end{equation}

\noindent where $\lambda$ is the observed wavelength, $A$ the area of a station, $B = 10 \, \rm{MHz}$ is the bandwidth, $x$ the comoving distance to the observed redshift, $y$ the depth of field.

The other main source of stochasticity in this problem is naturally the thermal noise of the instrument. The detector noise covariance matrix is

\begin{equation}\label{eq:ska_noise}
     C_{N}(k_i, k_j) \approx     \left(  \frac{\lambda^2 B T_{sys}}{A } \right)^2 \frac{\delta_{ij}}{B t}
\end{equation}

\noindent where $T_{sys}$ is the system temperature and $t$ the observation time. The standard deviation of the power spectrum caused by SKA thermal noise and CV is computed using a total covariance matrix $C = C_N + C_{CV}$ as \cite{Mcquinn} : 

\begin{equation}\label{eq:tot_noise}
    \delta P_{21}(k,z) = \left[\sum_{|k| = k}\left(  \frac{1}{ \frac{Ax^2 y }{\lambda  (z)^2 B^2} C(\textbf{k},\textbf{k})   } \right)^2 \right]^{-1/2}
\end{equation}

Technically, the Monte Carlo radiative transfer scheme in our simulation setup introduces an additional source of noise, however when using a reasonable number of photons (that depends on the numerical setup), it becomes negligible compared to the effect of the instrument. It must also be noted that, when we model a signal, the thermal noise and CV pose practical challenges of wildly different difficulties : at a given  astrophysical parameter set $\theta$, generating then adding realizations of the thermal noise to a noiseless 21-cm signal is extremely fast, but exploring the range of results offered by cosmic variance requires running many simulations with different density field initial conditions. Due to computation and storage limitations, \textsc{Loreli} II contains only one such simulation at each point in parameter space.

\subsection{The 21-cm power spectrum likelihood}

The stochastic processes are critical in the inference problem, as they mean that the same observed spectrum can be generated by different, albeit close, points in parameter space, with different probabilities. The sources of stochasticity shape the likelihood and therefore the posterior. 

A usual assumption is that the form of the likelihood is Gaussian :

\begin{equation}\label{eq:nondiag_likelihood}
    logL(y | \theta) = -  \frac{1}{2} \left(P_{21} - y(\theta) \right)^T \Sigma^{-1} \left(P_{21} - y(\theta )\right)
\end{equation}

\noindent where $y(\theta)$ is a 21-cm power spectrum generated at $\theta$ and $P_{21}$ the inference target.  $\Sigma$ is the covariance matrix : its diagonal represents the variance due to the aforementioned stochastic processes, and the off-diagonal terms imply coupling between modes induced by the complex physics of the 21-cm {(for instance the non-linear evolution of the density field, as well as heating and ionization state of the IGM)}. Nonetheless, the covariance is sometimes assumed to be fully diagonal, yielding 

\begin{equation}\label{eq:diag_likelihood}
    logL(y | \theta) = - \sum_{k,z} \frac{1}{2} \left( \frac{P_{21}(k,z) - y(\theta, k ,z )}{\sigma_{tot}(k,z)} \right)^2
\end{equation}

\noindent where $\sigma_{tot}$ contains the diagonal of $\Sigma$. This last expression, although inaccurate, is convenient and easily evaluated, which makes it widely used \citep[see][for a discussion]{Prelogovic2023}. 

In order to perform an accurate parameter inference using the \textsc{Loreli} simulations, two issues remain :

\begin{itemize}
    \item One must be able to evaluate the likelihood at more points than the 10 000 simulated parameter sets of \textsc{Loreli} II. Typical classical inference requires $\sim 10^{5-6}$ realizations for a 5D parameter space \citep{Prelogovic2023, TheHERACollaboration2022}. Some form of interpolation between the \textsc{Loreli} simulations is required.
    \item One would benefit from relaxing the Gaussian Likelihood approximation. To do so, one must find a way to evaluate the true, unknown likelihood of the signal.
\end{itemize}

\section{Inference methods}

The previous sections laid the groundwork for the machine learning methods we use to perform the EoR parameter inference. In this section, we present three approaches that we apply to the 21-cm power spectrum. The first consists in emulating the power spectrum with a neural network and using the emulator in a classical MCMC framework, the second is a direct prediction of the posterior distribution of the astrophysical parameters with Bayesian Neural Networks, and the last works by fitting the likelihood of our data with a Neural Density Estimator. All implementations rely on Tensorflow\footnote{\url{https://www.tensorflow.org/}} and the Keras\footnote{\url{https://keras.io/}} framework.
{In this section, we present and evaluate each inference method ; we compare them in the next section.}

\subsection{ Classical MCMC inference with an emulator}

\subsubsection{LorEMU}


\begin{figure*}
    \centering
    \includegraphics[scale = 0.8]{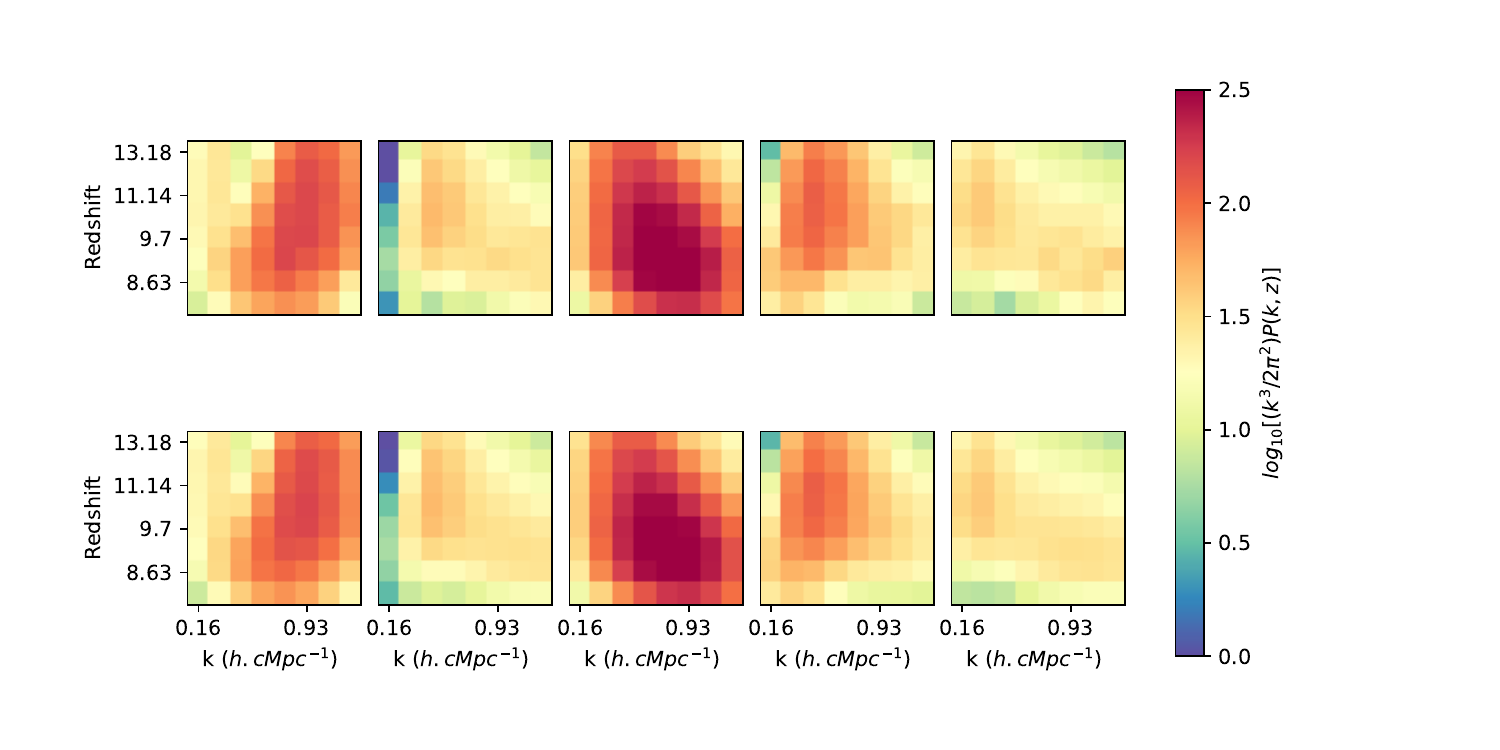}
    \caption{\textit{Top : } randomly selected emulated power spectra. \textit{Bottom : } the corresponding spectra from \textsc{Loreli} II.  {This figure is a simple illustration of the performances of LorEMU. }   }
    \label{fig:emu_mozaic}
\end{figure*}


        

    
    

\begin{table}[]
    \centering
    \begin{tabular}{|c|c|c|}
        \hline 
         \textbf{Layer type} & \textbf{Activation function}  \\
         \hline             

         Dense , 1024 neurons & Leaky ReLu ($\alpha = 0.1$)  \\
        Dense, 1024 neurons & Leaky ReLu ($\alpha = 0.1$)   \\
        Dense, 1024 neurons & Leaky ReLu ($\alpha = 0.1$) \\
        Dense, 64 neuron & linear   \\
        
    \end{tabular}
    \begin{tabular}{|c|c|c|}

    \hline
    \textbf{Optimizer} &  \textbf{Loss function} \\ 
    \hline
    RMSprop (learning rate = $10^{-3})$ & Mean Squared Error \\ 
    
    \hline 
    \end{tabular}
    
    \caption{Architecture and hyperparameters of \textsc{LorEMU}.  }
    \label{tab:archi_emu}
\end{table}

\begin{figure*}[h!]
    \centering
    \includegraphics[scale = 0.75]{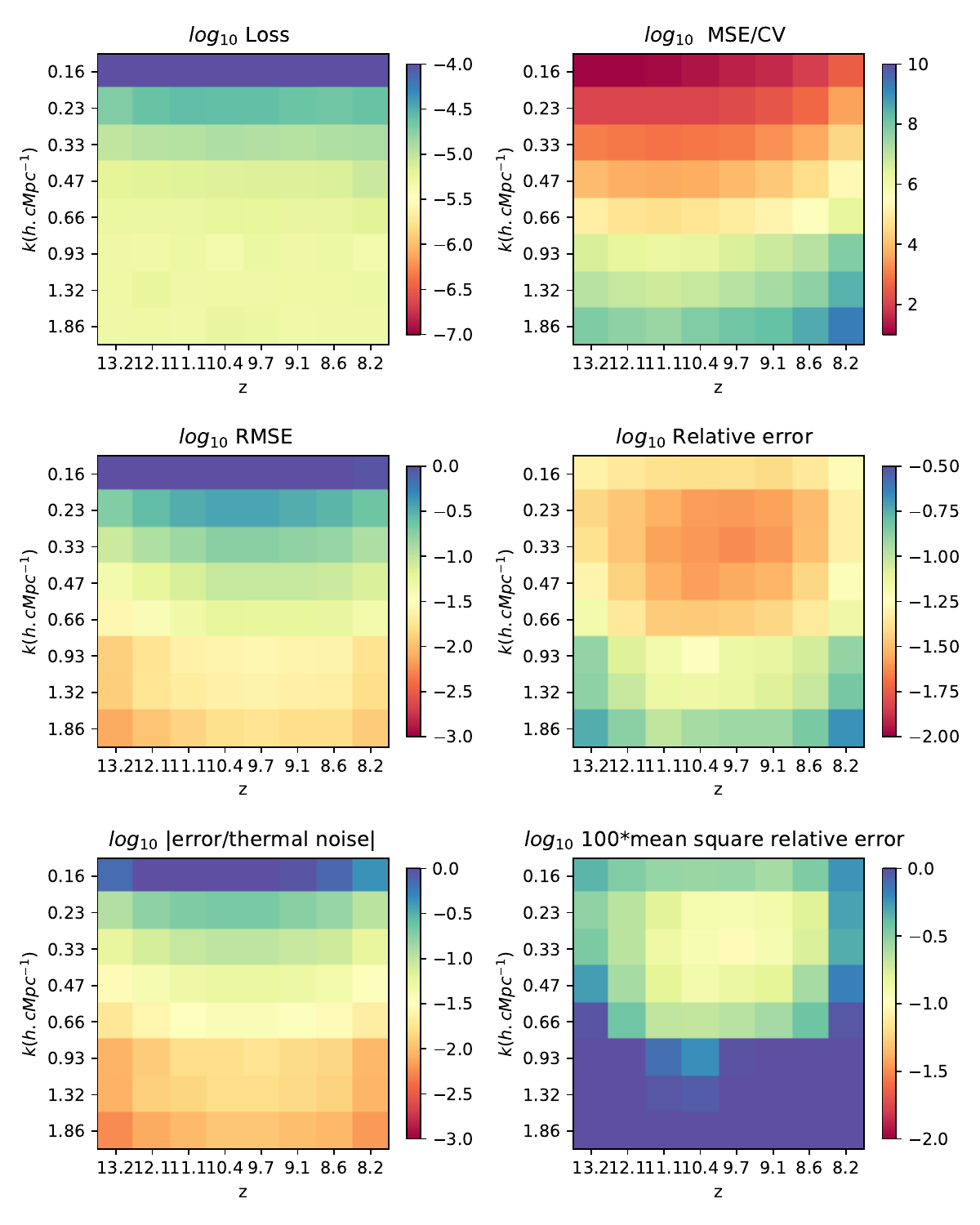}
    \caption{Performance metrics of an instance of LorEMU, averaged over the validation set. The network perform betters where CV and the thermal noise are weak. See text for details. }
    \label{fig:emu_error}
\end{figure*}

\begin{figure*}[h!]
    \centering
    \includegraphics[scale = 0.75]{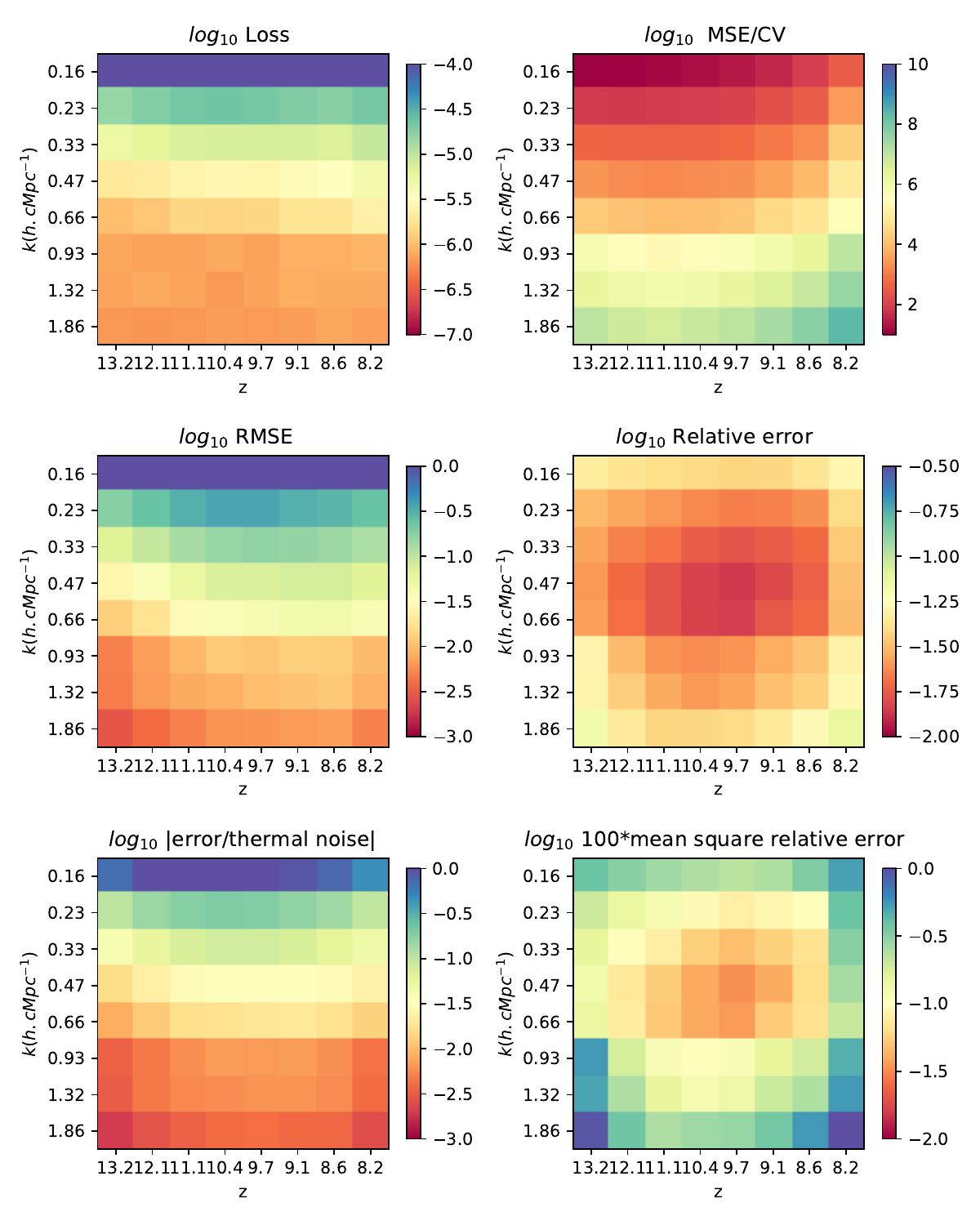}
    \caption{Similar to \ref{fig:emu_error}, but for the average predictions of 10 independently trained instances of LorEMU, leading to a significant performance improvement. For instance, the relative error typically decreases by a factor $3$.    }
    \label{fig:emu_error_stack}
\end{figure*}


The emulator, named LorEMU II, is a network trained to take as inputs the astrophysical parameters of the \textsc{Loreli} II simulations and output the corresponding power spectra. The full architecture is shown in table \ref{tab:archi_emu}. It was trained on $8\times8$ power spectra, computed from coeval 21-cm signal cubes at redshift $z =$ [13.18, 12.06, 11.14, 10.37,  9.7 ,  9.13,  8.63,  8.18] and wavenumbers $k =$ [0.16, 0.23, 0.33, 0.47, 0.66, 0.93 , 1.32 , 1.86] $\rm{h/cMpc}$ (thus we perform a multi-redshift inference).  The training set comprised $90\%$ of the \textsc{Loreli} II spectra while the rest formed the validation set. The logarithm of $\tau_{SF}$, $M_{min}$ and $f_X$ is taken, and all input parameters are normalized between $0$ and $1$. The first version of LorEMU, trained on \textsc{Loreli} I, emulated the logarithm of the spectra. It was later found that training on the square root of the signal-to-noise ratio leads to more accurate results : 
\begin{equation}\label{eq:norm_emu}
    \Tilde{P}_{21} = \sqrt{\frac{P_{21}}{\sigma_{thermal}}}
\end{equation}

\noindent where $\sigma_{thermal}$ is the standard deviation of the SKA thermal noise for 100h of observations, given by equations \ref{eq:ska_noise} and \ref{eq:tot_noise}. This normalization forces the network to learn the features that are observable by the instrument, and ignore those that are masked by the noise. This is not necessarily the case if the emulator simply predicts the spectra, as it may minimize its cost function by improving its performances in regions that would turn out to have a low signal-to-noise and will not be relevant during the inference. Additionally, the square root confines the signal-to-noise to a manageable dynamic range, {leading in this case to better results than a log}. \\

The training aimed at minimizing a Mean-Square Error (MSE) loss with an RMSprop optimizer using a learning rate of $10^{-3}$ and convergence was typically reached after $500$ epochs. 
Fig.  \ref{fig:emu_mozaic} shows a few randomly selected examples of simulated and emulated spectra, to qualitatively assess the quality of the results.\\

Quantitatively, Fig. \ref{fig:emu_error} shows different metrics to evaluate the performance of the network at each (k,z) bin, averaged over the validation set. All quantities, except for the loss, have been computed on the signal-to-noise $\frac{P_{21}}{\sigma_{thermal}}$ {(i.e. on de-normalized signals, not on $\Tilde{P}_{21}$ of Eq. \eqref{eq:norm_emu})}.  

The top-left plot shows the MSE. It exhibits a strong dependence in $k$ but not in $z$. The loss is the highest at large scales, where the cosmic variance is significant. This is detailed on the top-right plot, which shows the ratio of MSE and CV (for denormalized signals). {At large scales, the ratio converges to $\sim 1$, implying that the dominant factor of error is CV, and becomes negligible at small scales, where CV vanishes.} Ideally, we want the emulator to learn the CV-averaged power spectrum at each point in the parameter space. In practice, our loss shows no sign of overfitting, and we observed poor performance at large scales on the training set, which indicates that the network does not learn the specific CV realization that affects each signal in the dataset. Instead, it is hypothesized that it learns to average over the contribution of CV by smoothing out local fluctuations between points in the parameter space due to CV.

The loss and the Root-Mean-Square-Error (RMSE) are minimal at high $k$, as indicated on the centre-left plot. However, the relative error at these scales is at its worst, as shown on the centre-right plot. These are regions in $(k,z)$ where the thermal noise of the instrument is high, and the signal-to-noise consistently very low and easy to learn for the network. In regions of lower noise, where the emulator must be accurate in order to provide good inference results, the relative error shown on the mid-right plot is only a few $\%$ on average. {Since we train the network to predict a spectrum unaffected by the thermal noise, the error can indeed become small compared to the noise. However, we cannot train it to predict spectra unaffected by CV because we have only one realization at each parameter space point, and thus the error cannot become negligible compared to CV.} Also, we have not observed any clear systematic dependence of the error level on the astrophysical parameters. The bottom-left plot of Fig. \ref{fig:emu_error} shows the ratio of the absolute error and the thermal noise, while the bottom-right plot shows the mean square relative error. They confirm that the networks consistently perform best in the region most relevant for the inference. \\

Different metrics are exposed on figures \ref{fig:emu_error}, and they focus on different regions of $(k,z)$, but in the $0.1 < k < 0.9 \, {\rm{h.cMpc^{-1}}} $ range (i.e. in the low S/N region), the performances of LorEMU II are comparable with those of LorEMU I (trained on \textsc{Loreli} I, and studied in \cite{Meriot2024}, in terms of mean square relative error (which describes a variance), and they typically outperform the emulator presented in  \cite{Jennings2019} by a factor of $\sim 2 $.

In summary, the network performs best where the cosmic variance and the thermal noise is low. The relatively poor performances where there is noise and CV should be of minor consequence for the inferences, as this means that the network introduces errors where the sources of stochasticity already erase information about the physics of Reionization and not where this information is available and clean.

\subsubsection{Inference on mock data}

\begin{figure*}
    \centering
    \includegraphics[scale = 0.72]{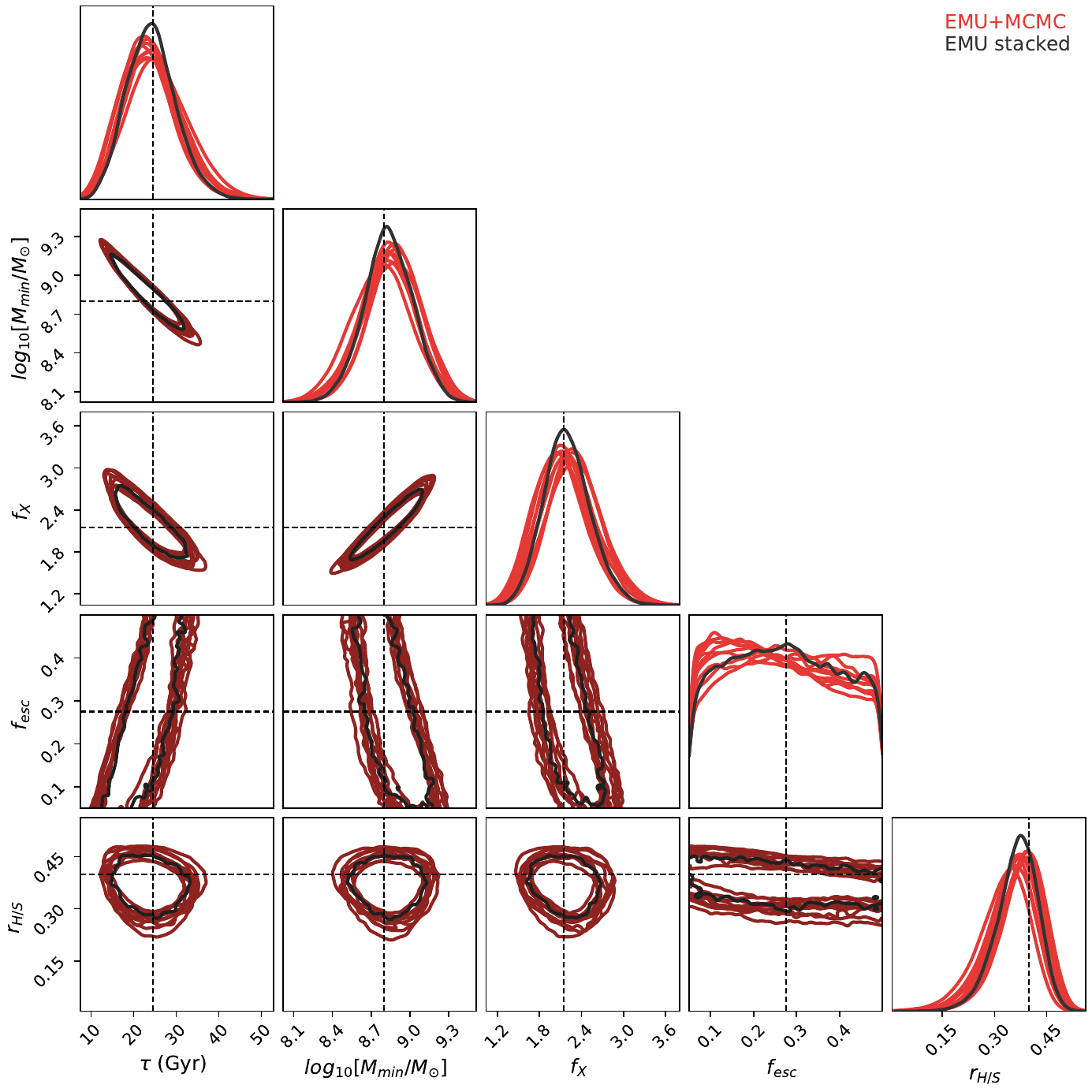}
    \caption{Results of inference on a noiseless \textsc{Loreli} II power spectrum using 10 independently trained versions of the LorEMU II emulator (\textit{red)}. The emulators show non-negligible variance. This uncertainty can be mitigated by averaging the predictions of the different instances of LorEMU during the inference (\textit{black}). {On 2D posteriors, only the $1\sigma$ contours are shown, for clarity.} }
    \label{fig:emu_inf}
\end{figure*}


LorEMU II can generate an $8\times8$ power spectrum in a few $ms$ on an Nvidia T600 mobile GPU and can therefore reasonably be used in an MCMC pipeline. Let us perform a series of inferences on artificial data : a \textsc{Loreli} II signal, to which we can add a thermal noise realization corresponding to 100h of SKA observations.  We assume that the likelihood is given by Eq. \ref{eq:diag_likelihood}. This assumption is challenged by e.g. \cite{Prelogovic2023}, but this paper also shows that the off diagonal terms are mostly erased by the presence of thermal noise of 1000h of observations, significantly weaker than in our setup. We now need to specify the $\sigma_{tot}$ that appears in the likelihood : 

\begin{equation}
        \sigma_{tot} = \sqrt{\sigma_{thermal}^2 + \sigma_{CV}^2 + \sigma_{train}^2}.
    \end{equation}
\noindent
$\sigma_{thermal}$ and $\sigma_{CV}$ are given by equations \ref{eq:cv_noise} to \ref{eq:tot_noise} and {contain the standard deviation associated with} the thermal noise and CV, respectively. The last term, $\sigma_{train}$, is caused by the imperfections of the emulator : for instance, the prediction of LorEMU depends on the random initialization of its weights before training.
Since the {MSE} of LorEMU is non-zero and does not {show a strong dependency with the} parameters $\theta$, we follow the procedure outlined in \cite{Meriot2024} and consider that the emulated signal $ P_{21}^{(emu)}$ is the corresponding simulated signal $ P_{21}^{(simu)}$ affected by a training noise i.e. a random perturbation of standard deviation $\sigma_{train}$, given at each $\theta$ by :

\begin{equation}\label{eq:sig_train}
  \sigma_{train}^2(k,z,\mathbf{\theta}) = \frac{1}{N}  \sum_N \left( P_{21}^{(emu)}(k,z,\mathbf{\theta}) - P_{21}^{(simu)}(k,z,\mathbf{\theta}) \right)^2  
\end{equation}

\noindent where $N$ is a large number of emulators trained with different initializations. Since $P_{21}^{(simu)}$ is only available at the parameter sets of \textsc{Loreli} II and not for any $\mathbf{\theta}$, {and since the MSE of LorEMU does not strongly depend on $\mathbf{\theta}$, } we estimate $\sigma_{train}$ by averaging this quantity over $\mathbf{\theta}$. 

Figure \ref{fig:emu_error_stack} shows the performance of a "stacked emulator", i.e. measured using the average predictions of 10 different instances of LorEMU, trained with different weights initialization and on different splits of the data to constitute the training and validation sets. Stacking the emulators results in an improvement of the performance and a decrease of the RMSE  by a factor $\sqrt{10} \sim 3$. {This is consistent with our hypothesis that the training error behaves as a random noise of mean zero, as it is the result expected from the central limit theorem.}  The relative absolute error of the stacked emulator is smaller than $5\%$ across almost all $k$ and $z$ (when averaged over $\theta$) and the absolute error is consistently significantly below the level of the thermal noise {($\lesssim 5 \%$ for $k \gtrsim0.4 \, {\rm{h.Mpc^{-1}}}$)}, as shown on the bottom-left panel. The average relative error consistently decreased compared to a single instance of LorEMU to reach $3\%$ on average, and displays a smaller variance in the performances.

For an inference target, we use a simulated signal from \textsc{Loreli} II with the following parameters : $M_{min} = 10^{8.8} \Msun$, $\tau_{SF} = 25 $ Gyr, $f_X = 2.15$,  $r_{H/S} = 0.4$, $f_{esc} = 0.275$, near the centre of the \textsc{Loreli} II parameter space (our prior). In this section, we perform the inferences on a target signal unaffected by noise but with the effect of stochastic processes duly included in the inference frameworks. {Since the signal is not affected by a noise realization, we expect the posterior to be well centred around the true value of the parameters because we are in a situation where stochastic processes are mostly Gaussian. Note however that, in the general case, a multi-dimensional posterior with a complex shape does not necessarily result in 1-D marginalized posteriors centred on the true parameter values, even for a noiseless signal. The assessment of the results of the inferences in this section is heuristic, a formal validation is the object of the next section. Moreover, the target signal is still affected by a CV realization, which may lead to a small bias in the posterior. Of course, an imperfect inference framework (emulator training in this case) should produce biases among other errors. 

We run the MCMC with each of the 10 instances of LorEMU individually first, and then using the stacked emulator. The results are shown on Fig. \ref{fig:emu_inf}. The resulting inference exhibits some variation depending on the actual emulator used, but the true values of the parameters always lie within the 68$\%$ confidence interval and the remaining bias is small compared to the $1\sigma$ {(remember that here the target is a  signal not affected by thermal noise, the largest source of stochasticity, but thermal does contribute to the posterior variance)}. Interestingly, all parameters are well constrained except for the escape fraction, which is common across all methods. It is possible that including the values of the power spectra at lower redshifts would allow better constraints.

For comparison, figure 12 in \cite{Meriot2024} shows similar inferences performed using LorEMU I, on a signal from \textsc{Loreli} I. Qualitatively, the inferences on Fig. \ref{fig:emu_inf} done with \textsc{Loreli} II  seem to exhibit less variance than those done with \textsc{Loreli} I. The inference obtained with the average prediction of 10 instances of LorEMU I seems also significantly more biased than what we observe with the average LorEMU II. This suggests that the architectural modifications to LorEMU II and the better sampling of \textsc{Loreli} II lead to better results compared to the work presented in \cite{Meriot2024}.

Importantly, the posterior contains several points of the \textsc{Loreli} II grid within the $99\%$ confidence regions. While not critical for the emulator, the other networks architectures described in the following need to train on signal affected by an instrumental noise realization. Ensuring the sampling of the parameter space is dense enough to cause some overlap between the noised signal distributions of neighbouring points of the grid may help to avoid a pathological regime where the networks learn to affect every noised versions of the same un-noised signal to the exact same parameter set with a very high probability instead of properly yielding the posterior or the likelihood. In other words, it should prevent the networks from becoming absurdly overconfident. The design of the \textsc{Loreli} II grid was motivated by the results obtained with \textsc{Loreli} I, and this inference tends to confirm that the database is suited for an inference on data affected by a noise at least as strong as the one on 100h of SKA observations.

\subsection{Simulation Based Inference}


In addition to the emulator, we explore in this work the use of simulation-based inference (SBI), also called likelihood-free inference. SBI relies on a simple fact : when repeatedly taking a given input $\theta$ (in our case, the astrophysical parameters), a simulation code will produce a distribution of outputs $D$. If the code is deterministic, this distribution is a Dirac, but in our case, the relevant output is  a realistic mock observed 21-cm power spectrum which is affected by at least two sources of stochasticity : the thermal noise of the instrument and the cosmic variance. By definition, these outputs are drawn from $P(D|\theta)$, a likelihood function inherent in the simulation pipeline. If this likelihood function can be fitted and quickly evaluated, it can be used in a MCMC algorithm. An alternative is to directly fit the posterior. Different network architectures can also be used to fit the desired distribution.

As discussed in e.g. \cite{Prelogovic2024}, using a Gaussian shape for the likelihood of the 21-cm power spectrum is an approximation, and this family of methods is in principle a way to relax it. For different variations and evaluations of their performance, see for instance \cite{Prelogovic2023}, \cite{Zhao2022b}, \cite{Saxena2023}. In the following, we consider two approaches : we fit the posterior using a Bayesian Neural Network and fit the likelihood using a Neural Density Estimator.

\subsection{ SBI with Bayesian Neural Networks : posterior estimation}


First, we study how to directly predict the posterior distribution of the astrophysical parameters using a Bayesian Neural Network. These networks contain so-called \textit{variational layers} which differs from classic layers in the fact that the connections between neurons are not parametrized by single values for the weights but rather by weight distributions. When the layer is called, the input of the neurons are weighted by values drawn from these distributions. BNNs are probabilistic : repeatedly feeding them the same input will result in a distribution of outputs. For a review, the reader is redirected to \cite{Gal2016}, \cite{Jospin2022}. Note that \cite{Hortua} presented a first, brief application (using dropout layers, not true weights distributions) of BNNs on the 21-cm signal. In the following, we expand on their work, in particular in terms of evaluation of the performance of the network. Let us first briefly summarize how BNNs work.

\subsubsection{Variational inference}

\begin{figure*}[h]
    \centering
    \includegraphics[scale = 0.85]{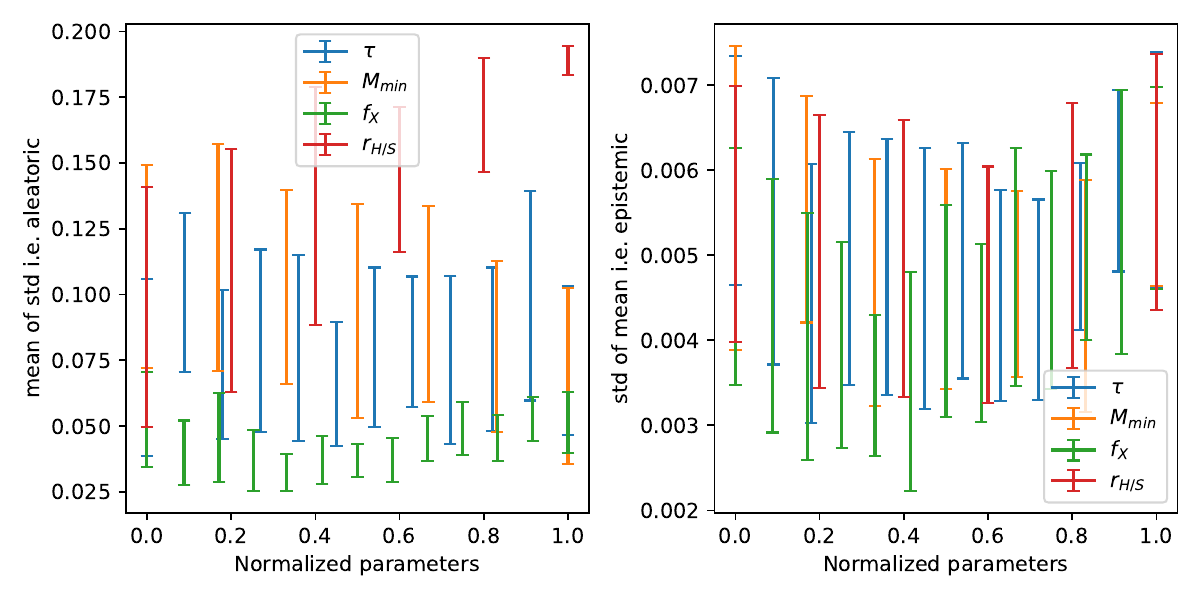}
    \caption{\textit{Left} : average standard deviation of the predicted distributions of each parameter by the BNN. This quantifies the aleatoric uncertainty, i.e. the uncertainty caused by the noise in the data. \textit{Right :} Standard deviation of the predicted means. This is the epistemic uncertainty, i.e. the error caused by the sparsity of the training sample. It is an order of magnitude smaller than the aleatoric uncertainty, indicating that the instrumental noise is the dominant source of variance.}
    \label{fig:bnn_errs}
\end{figure*}

\begin{figure}[h]
    \centering
    \includegraphics[scale = 0.9]{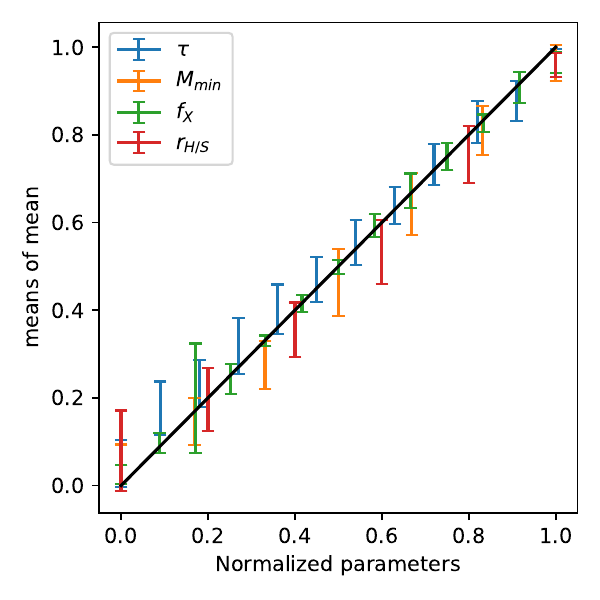}
    \caption{Means of: average standard deviation of the predicted distributions of each parameter by the BNN. This quantifies the aleatoric uncertainty, i.e. the predicted distributions by the BNN compared to the true parameters, computed on the validation set. While the predicted means are very stable (as shown on the right panel of \ref{fig:bnn_errs}), the network shows a significant bias for all parameters save for $f_{X}$ in $\sim 10\%$ of the cases.  }
    \label{fig:bnn_means}
\end{figure}

 During training, given a dataset $D = {(x,y)}$, one wishes for the parameters of the network to be adjusted so that the networks responds to an input $x$ (in our case, a 21-cm power spectrum affected by a mock realization of SKA noise) with an output close to $y$ (the corresponding astrophysical parameters) with a high probability. That is, given some prior $p(\rm{\omega})$ set over the weights $\rm{\omega}$ , one wishes to determine $p(\rm{\omega} | D)$, the posterior distribution of the network's weights given some data $D$. To do so, we naturally use Bayes theorem : 

 \begin{equation}
     p({\rm{\omega}} | D) = \frac{p({\rm{\omega}}) p(D|{\rm{\omega}}) }{p(D)}
 \end{equation}

 \noindent where $p(D) = \int p(D|\rm{\omega}) \rm{d\omega}$ is the evidence. Calculating this integral (i.e. marginalizing over the weights) is intractable except in the simplest cases. In practice, one turns to an approximation : the true posteriors $p(\rm{\omega} | D) $ are replaced by parametrized distributions $q_{\phi}$,  characterized by some numerical parameters $\phi$. A common choice, adopted in the following, is to set Gaussian distributions over the weights. As a note, classic layers with deterministic weights can be perceived as Dirac distributions in this formalism. To ensure that the $q_{\phi}$ are good approximations of the posterior, the $\phi$ parameters are tuned to minimize the Kullback-Leibler (KL) divergence between the true and approximated posteriors :

 \begin{equation}
     KL(q_{\phi}({\rm{\omega}}), p({\rm{\omega}} | D)  ) = \int q_{\phi}({\rm{\omega}})   log \frac{q_{\phi}({\rm{\omega}})}{p({\rm{\omega}} | D)  } {\rm{d\omega}}
 \end{equation}

This is equivalent to maximizing the Evidence Lower Bound \citep[ELBO][]{Gal2016, Jospin2022}  : 

\begin{equation}\label{eq:elbo}
    L_{ELBO} =  \int q_{\phi}({\rm{\omega}})  log ({p(D|{\rm{\omega}}))  } {\rm{d\omega}} - KL( q_{\phi}({\rm{\omega}}) , p({\rm{\omega}}) ).
\end{equation}

\begin{table*}[]
    \centering
    \begin{tabular}{|c|c|c|}
        \hline 
         \textbf{Layer type} & \textbf{Activation function}  \\
         \hline             

         Dense , 1024 neurons & Leaky ReLu ($\alpha = 0.2$)  \\
        Dense, 1024 neurons & Leaky ReLu ($\alpha = 0.2$)   \\
        Dense, 1024 neurons & Leaky ReLu ($\alpha = 0.2$) \\
        Dense Variational, $2 \times (5  + 25)$ neurons  & Linear   \\
                \hline 

    \end{tabular}
    \begin{tabular}{|c|c|c|}

    \hline
    \textbf{Optimizer} &  \textbf{Loss function} & \textbf{Batch size} \\
    \hline
    RMSprop (learning rate = $5\times10^{-4})$ &Log prob & 32\\
    
    \hline 
    \end{tabular}
    
    \caption{Architecture and hyperparameters of the BNN. The neurons of the last layer predict the parameters of a mixture of two multivariate normal distributions.   }
    \label{tab:archi_bnn}
\end{table*}

Maximizing the first term means adjusting  $\phi$ to maximize the likelihood, while the second terms act as a regularization term : it penalizes complex $q_{\phi}$. This defines a minimization objective : $-L_{ELBO}$. 
In theory, during a single epoch, in order to estimate $-L_{ELBO}$, a single example must be passed many times to properly sample the integrand. In practice, it is estimated by a single forward pass \citep{Gal2015}. This adds another source of noise to the training, but is assumed to converge properly in the limit of a large dataset. {This leads to a discrete version of Eq. \eqref{eq:elbo} :}

    \begin{equation}\label{eq:dist}
    L_{ELBO} =  \sum_i  q_{\phi}({\rm{\omega}}_i)  log ({p(D_i|{\rm{\omega}}_i))  } - KL( q_{\phi}({\rm{\omega}}_i) , p({\rm{\omega}}_i) )
\end{equation}

{ \noindent where the sum is taken over the elements of the dataset. For each of them, a set of weights $\omega_i$ is sampled from the layer distributions. }



 
An additional feature of the BNN lies at its last layer. The BNN does not predict single values for the astrophysical parameters we wish to infer from the data. Instead, it predicts parameters of distributions. We chose to predict a mean vector $\mu$ and the full covariance matrix  $C$ that will capture the correlations between the astrophysical parameters. For instance, with one Gaussian component, the {probability of the network outputting $y$ given an input $x$ and  a set of weight values $\omega$ (sampled from the parametrized distributions sets over the weights) can be written :

 \begin{equation}
  -log p(y|x,\omega) = \frac{1}{2}log( (2\pi)^d |C(\omega)| ) +  \frac{1}{2} (y - \mu(\omega))^T C^{-1} (y - \mu(\omega)).
 \end{equation}



{This log-probability $log p(y|x,\omega)$ is the  $log p(D |\omega)$  that appears in the loss function of Eq. $\eqref{eq:elbo}$.    
}

\subsubsection{{Uncertainties with BNNs}}

While the emulator-based approach requires an explicit prescription of each of the variances induced by all stochastic processes at play, the BNN does not, and can instead be used to produce an estimate of two forms of variance of interest. Indeed, after passing the same input $x$  to the network $N$ times, the model predicts a set of means $\{ \mu_n\}_{1\leq n \leq N}$ and covariance matrices  $\{ C_n\}_{1\leq n \leq N}$. This allows us to compute an estimate of the average prediction

\begin{equation}
    \langle \mu \rangle = \frac{1}{N}\sum_{n=1}^N \mu_n
\end{equation}

\noindent as well as an estimate of the covariance of the model 

\begin{equation}
    \langle C \rangle = \frac{1}{N}\sum_{n=1}^N C_n + \frac{1}{N}\sum_{n=1}^N (\mu -  \langle \mu \rangle)(\mu -  \langle \mu \rangle)^T.
\end{equation}

In the last equation, the first sum characterizes the \textit{aleatoric uncertainty}, while the second sum is a measure of the \textit{epistemic uncertainty}. The former is the uncertainty that arises from the noise inherent in the data (in this case, the instrumental noise and the cosmic variance), and the latter captures the variance caused by an imperfect training of the network, for instance due to a limited density of the training sample or insufficient optimization. In this sense, what the epistemic uncertainty represents may be similar to the model error introduced in the explicit likelihood in the previous section.

    The data on which the network trains is the power spectrum noised by instrumental effects, given in section \ref{sec:stoch}. To produce one realisation of the noise, one approach would be to draw values from a Gaussian distribution of mean $0$ and a standard deviation given by Eq. \ref{eq:tot_noise}. While this works in cases where the signal-to-noise is high, it leads to negative values where the noise is strong. These negative values are not physical, but would not impede the training if it were not for the need of normalization. Indeed, the data must be normalized to an interval close to $[0, 1]$ while maintaining the dynamic range i.e. having a network that learns the features of weak spectra just as well as strong ones. The negative values prevent the use of Eq. \ref{eq:norm_emu}, or a logarithm. S-shaped functions like hyperbolic tangents or sigmoids may divide a set consisting of one power spectrum at one $k,z$ affected by different noise realizations in two clusters depending on whether the noise added or subtracted power (i.e. it tends to create bimodal likelihoods). The network must then learn that these two clusters characterize the same data, which has empirically proven to be a challenge. A first solution is to cut negative values and replace them with some floor value such as 0, another is to compute the power spectra of noised $\delta T_b$ cubes and perform the inference on the power spectrum that includes the contribution of the  noise. The latter would produce physically accurate results but is much more computationally intensive than the former. For now, we only use the first method.


\begin{figure*}
    \centering
        \includegraphics[scale = 0.75]{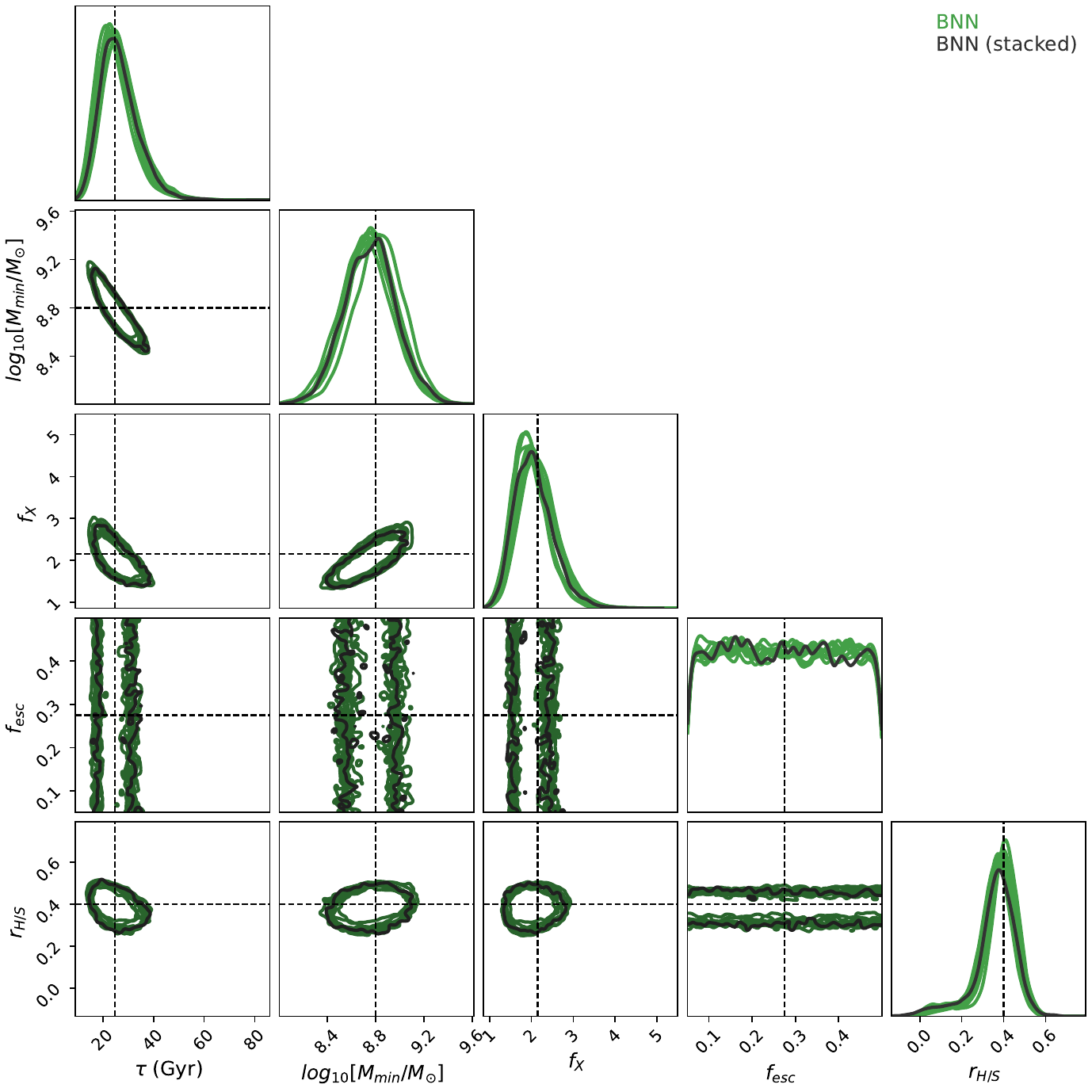}
    \caption{Posteriors estimated by repeatedly forward passing a noiseless \textsc{Licorice} power spectrum to 10 independently trained BNNs (\textit{green}), and to the stacked BNN (\textit{black}). The results show little variance, and seem to converge to slightly biased results. The posterior on $r_{H/S}$ shows a long tail not retrieved by the other inference methods.}
    \label{fig:bnn_inf}
\end{figure*}

    The architecture of the network is shown on table \ref{tab:archi_bnn}. While not heavier than the emulator, it is considerably harder to train, as the variational layers introduce a large amount of stochasticity  in the training. On the upside, it acts as regularization, and we never observed any sign of overfitting while training these networks. 

    Figure \ref{fig:bnn_errs} shows, for the different values of the parameters, the aleatoric (left) and epistemic (right) errors, i.e. the average variances and variances of predicted means, computed on the validation set (without any instrumental noise). The means are comparatively extremely stable, which tends to show that the training sample is dense enough to constrain the weights of the network, and adding more data should only marginally improve the result. The predicted variances, which capture the noise in the data (i.e. the effect of CV and the thermal noise) are typically an order of magnitude above the epistemic error, and again typically larger than the parameter grid step, which should enable the network to generalize to signals generated using parameters sets not in Loreli II (but still inside the volume of parameter space sampled in the database). Interestingly, the variance exhibits a clear correlation with $r_{H/S}$, {a trend we weakly observe with the other methods. It could stem from the fact that a high $r_{H/S}$ delays and homogenizes the heating, thus decreasing the signal-to-noise ratio}.
    
    Figure \ref{fig:bnn_means} shows the means of the parameter distributions {predicted by the BNN when applied to a noiseless signal} compared to the true values of the parameters. Most ($\sim 90, \%$ see discussion in Sect. \ref{sec:noiseless}) lie on the $y=x$ line. However, the network underperforms in many cases. First, the BNN tends to overestimate $r_{H/S}$ when the true value is 0, and tends to underestimate this parameter elsewhere. The network also has a tendency to underestimate $M_{min}$ and overestimate $\tau_{SF}$ (as it properly captures the correlation between the two), typically by $0.1$, which is close to the parameter grid step.  

    Like in the case of the emulator, we trained 10 instances of the BNN with different weight initialization. In this case, stacking the networks does not lead to a significant improvement of the results. The dispersion of the predictions decreases, but their average does not converge significantly toward the true values of the parameters. In other words, the error bars on Fig. \ref{fig:bnn_means} slightly decrease, but the points do not significantly get closer to the $y=x$ line. This is not unexpected, since probabilistic networks are often perceived as a way to marginalize over a distribution of networks. As such, these approaches are less sensitive to their weight initializations. However, this also means that the networks systematically converge during training towards weights that cause wrong predictions of the astrophysical parameters, as the errors displayed on e.g. Fig. \ref{fig:bnn_means} can be significant. This might suggest that the network is sensitive to cosmic variance or some unidentified outliers in the data. 
    
    Figure \ref{fig:bnn_inf} shows the predictions of the 10 BNNs after forward-passing a noiseless signal 10 000 times, as well as the average posterior. The maxima of likelihood are slightly offset, a systematic bias that is not corrected by the stacking. The posterior on $r_{H/S}$ appears slightly bimodal, a feature missing when using the other inference methods. {A comparison with the results of the other inference methods can be found in the next section. }


\subsection{  {SBI with Neural Density Estimators : likelihood fitting}  }



\begin{figure*}
    \centering
        \includegraphics[scale = 0.72]{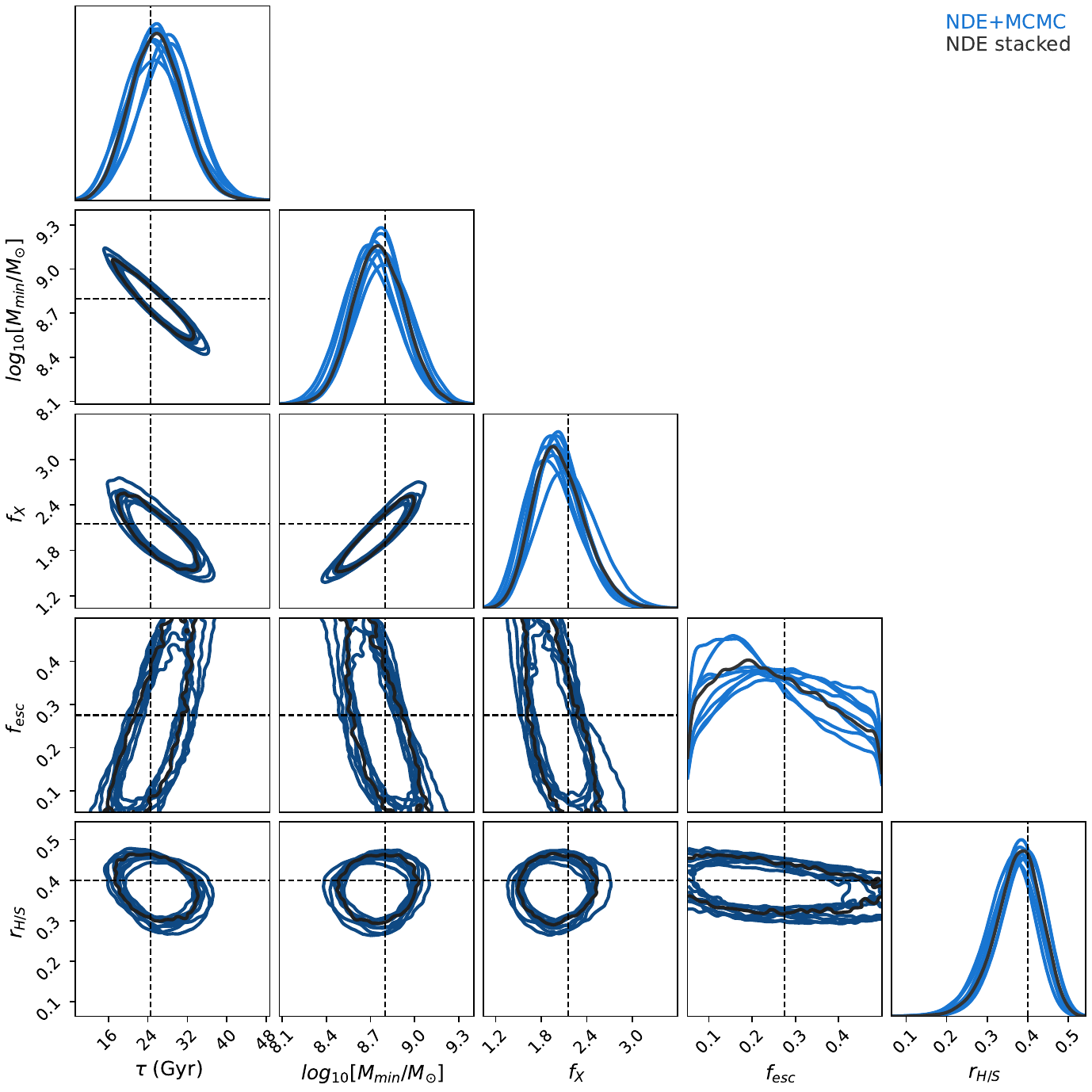}
    \caption{Similar to \ref{fig:emu_inf} but for 10 independently trained NDEs. They exhibit a similar amount of training variance as the emulators, but seem to display a larger systematic bias.  }
    \label{fig:nde_inf}
\end{figure*}

\begin{table*}[h]
    \centering
    \begin{tabular}{|c|c|c|}
        \hline 
         \textbf{Layer type} & \textbf{Activation function}  \\
         \hline             

         Dense , 64 neurons & tanh  \\
        Dense, 64 neurons & tanh   \\
        Dense, 64 neurons & tanh \\
        Dense, $4 \times 64$ neurons & tanh   \\
        (weights) Dense, 3 neurons & softmax  \\
        (means) Dense, $64\times3$ neurons & sigmoid  \\
        (variances) Dense,  $64\times3$ neurons & softplus  \\
        (off-diagonal terms) Dense, $64\times 4 \times 3$ neurons & linear  \\
                \hline 

    \end{tabular}
    \begin{tabular}{|c|c|c|}

    \hline
    \textbf{Optimizer} &  \textbf{Loss function} & \textbf{Batch size} \\
    \hline
    RMSprop (learning rate = $2\times10^{-3})$ &Log prob & 32\\
    
    \hline 
    \end{tabular}
    
    \caption{Architecture and hyperparameters of the NDE. The weights, means, variances, and off-diagonal terms are predicted by different output layers.  }
    \label{tab:archi_nde}
\end{table*}

We train a NDE to fit the distribution of the data $\{ D_n\}_{1\leq n \leq N}$ given the astrophysical parameters $\{ \theta_n\}_{1\leq n \leq N}$, i.e. the likelihood of our physical model. To do so, one must choose an arbitrary functional form $L_{NDE}(D|\theta)$ for the approximate likelihood, parametrized by the output of the NDE. The dataset to fit is the \textsc{Loreli} power spectra of the 21-cm signal, augmented by 100 realizations of the SKA thermal noise on the fly at each training epoch. As previously discussed, it would ideally contain several simulations per parameter set to sample cosmic variance, but this is not computationally feasible. Crucially, we use the same method to preprocess the spectra as with the Bayesian Neutral Networks.

\begin{figure*}
    \centering
    \includegraphics[scale = 0.72]{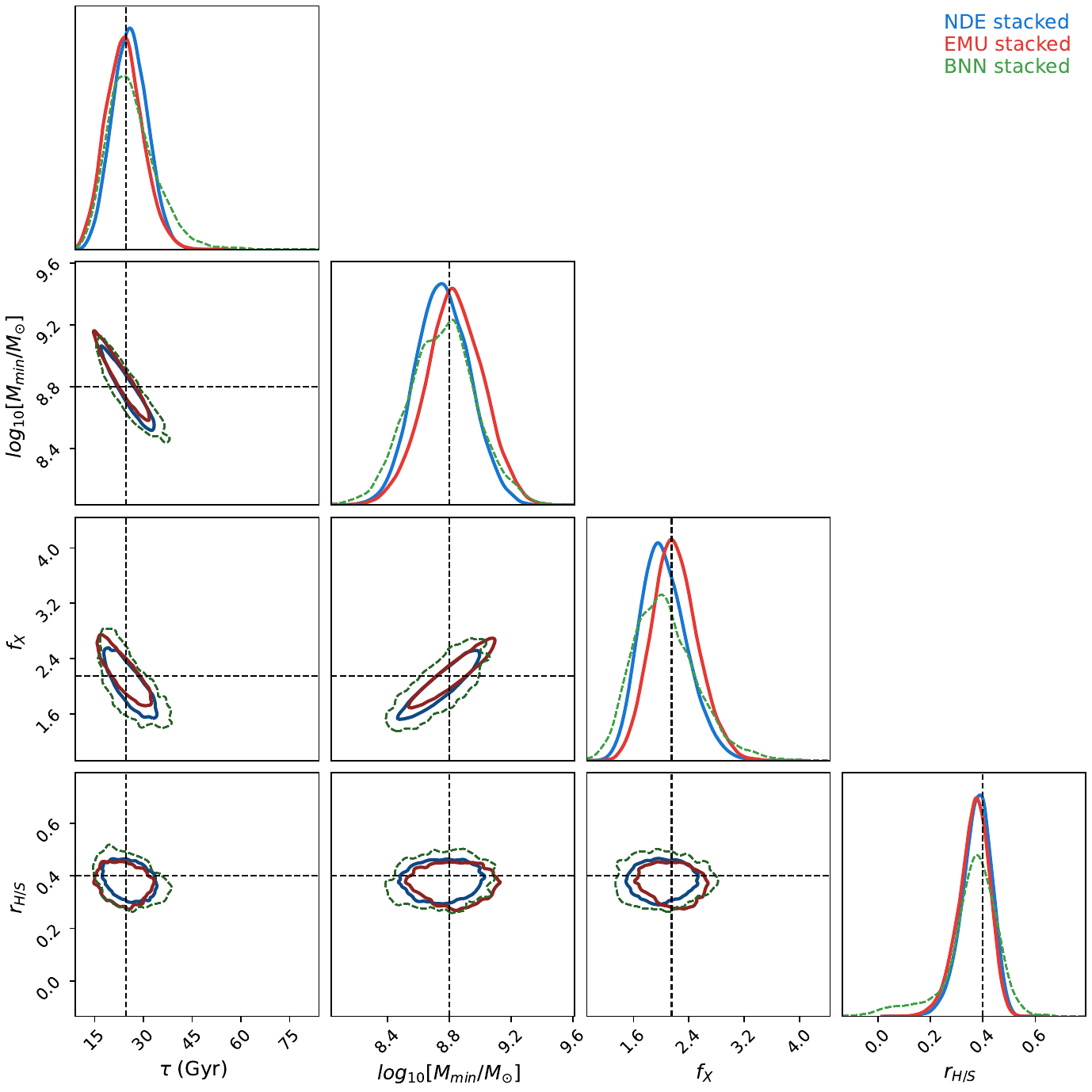}
    \caption{Inference with all (stacked) networks. All three results agree to a large extent. the maxima of likelihood are close : they are all with the $1\sigma$ of each other and within a grid step. Nonetheless, the BNN is systematically less confident than the others, while the predictions of NDE are slightly more biased. For clarity, we exclude $f_{esc}$ from the analysis, as it is systematically poorly reconstructed by all three methods.  }
    \label{fig:all_inf}
\end{figure*}

The network is trained so that the parametrized likelihood fits the true likelihood $L_{True}$ of the model. {Following \cite{Alsing2019}, in theory, we want to minimize the KL divergence between the parametrized likelihood of the NDE and the true likelihood,  }

\begin{equation}
  KL( L_{NDE}, L_{True}) =  \int L_{True}(D|\theta) \log \left( \frac{L_{True}(D|\theta)}{L_{NDE}(D|\theta)} \right) {\rm{d}}D .
\end{equation}
\noindent
{Minimizing this $KL$ is equivalent to minimizing $-\int L_{True}(D|\theta) \log \left( L_{NDE}(D|\theta) \right) {\rm{d}}D$, as we can only tune the parameters of $L_{NDE}$.} {In practice, we do not know $L_{True}$ but our training set is composed of draws in $L_{True}$, and minimizing the $KL$ divergence of the above equation is achieved  }  by minimizing the loss function, 

\begin{equation}
    loss_{NDE} = -\frac{1}{N} \sum_{n=1}^N  logL_{NDE}(D_n|\theta_n) .
\end{equation}

Once the network is trained to fit the true likelihood of the model, it can be put into an MCMC pipeline to perform the inference. Compared to the emulator-based method, this approach has the advantage of not requiring an explicit likelihood, it only requires a parametric functional form for the approximate likelihood that is flexible enough to fit the true likelihood with the desired accuracy. However, as is common with machine learning, a network made too flexible may be prone to overfitting. 
The functional form we choose for the parametrized likelihood is a multivariate Gaussian Mixture Model with $\eta = 3$ components, allowing the network to fit distributions more complex than the approximate Gaussian likelihood but with a manageable number of free parameters}. As shown in table \ref{tab:archi_nde}, the network consists of a simple multilayer perceptron with three layers of $64$ neurons each. Each parameter of the mixture model (i.e. the respective weights, means, diagonal terms and off diagonal terms of the Gaussian components) are predicted by an additional 64 neurons layer.

Training occurs over 35 epochs using an RMSprop optimizer and a learning rate of $2 \times 10^{-3}$, and the training and validation losses show no sign of overfitting.

As before, we illustrate the method by performing an inference on a \textit{noiseless} signal with 10 independently trained NDEs.
The results are shown on Fig. \ref{fig:nde_inf}. Again, we observe some variance of the maxima of likelihood, indicating a non-negligible training variance, but the true values of the parameters always lie well within the 68$\%$ confidence intervals.
As an attempt to reduce this variance, the figure also shows another inference performed while stacking the NDEs, i.e. by using their averaged predictions for the log-likelihoods. This stacked NDE still seems to converge to a solution that is not perfectly centred around the true value of the parameters, a behavior consistent with the one of the BNN. However, the dispersion of the result is reduced, and we will consider only the stacked NDE in the following.    


\section{Comparison between inference methods}\label{sec:comparison}

We presented the 3 inference methods used in this work and showed inferences performed on the same signal as an illustration. We now wish to identify which method to use to interpret an eventual detection of the 21-cm power spectrum. To do so, we need to evaluate 

\begin{itemize}
    \item their accuracy, i.e. whether the peak of the posterior distribution produced by an inference method is indeed the right one.  In an ideal and simple setting, {given that the noise is Gaussian distributed and of mean $0$}, inference on the average (equivalent to noiseless in this case) signal produced by a parameter set $\theta$ should yield a posterior that peaks at $\theta$, the true values of parameters. {This is true for simple cases and for a class of locally well-behaved models. We cannot guarantee that our model of the power spectrum belongs to that class, but the posteriors of the previous sections, as well as from other 21-cm inference works \citep[e.g.][]{Prelogovic2023, Zhao2022, Greig2018, Greig2015} do suggest that the posterior should peak at or near the true parameters.}  Effects like training variance and CV may create systematic bias and dispersion in the outcome of the inferences, {but we expect them to be small}. 
    \item their precision, i.e. we want to ensure that the posteriors we recover are not over- or underconfident. Following  \cite{Cook2006} and \cite{Talts2018}, we must check that posteriors obtained from noised signals stemming from $\theta$ should for instance contain $\theta$ 68\% of the time in the $68\%$ confidence contour, 95\% of the time in the 95\% confidence contour, etc. {The best method is not necessarily the one that yields the most confident posterior.}
\end{itemize}

\subsection{A heuristic approach : inference on noiseless signals}\label{sec:noiseless}

As a first element of comparison, Fig. \ref{fig:all_inf} shows inferences using all three methods (using stacked networks). On this signal, the BNN appears slightly less confident than the other methods, and the NDE shows the most bias. However, the differences between maxima of likelihood are of the order of $\sim 0.5 \sigma$, similar to what can be found in works such as \cite{Prelogovic2023}.

Naturally, a single inference is not enough to evaluate the performance of these methods. To provide a robust comparison, we perform inferences on $100$ randomly selected noiseless signals. For each inference, we find the maximum of each 1D marginalized posterior and compute the bias $\delta \mu$ (the difference between the true values of the parameters and the values parameters where the posterior peaks) and the standard deviation $\sigma$ of the 1D posteriors. Since what we expect from the networks is effectively to be able to interpolate between the signals of the database, a useful scale against which we can compare our error is the typical grid step of \textsc{Loreli} II, $\Delta_{grid} = {0.089,0.169, 0.088, 0.5, 0.2 }$ for the normalized $\tau_{SF}$, $M_{min}$, $f_X$, $r_{H/S}$ respectively.  However, since the escape fraction $f_{esc}$ is systematically very poorly constrained, we focus on the other four parameters. Note that the fact that our models cannot discriminate between different values $f_{esc}$ may force them to learn to average over cosmic variance {(at least to some extent)}, as they are presented with several power spectra affected by different realization of CV but labelled by the same set of the meaningful parameters.  \\

To evaluate the accuracy of our methods, we compare the position of the maximum of the 1D posterior and the true value of the astrophysical parameters for each inference on noiseless signals.  Distributions of $\frac{\delta \mu}{\Delta_{grid}}$ (the absolute bias) and  $\frac{\delta \mu}{\sigma}$ (the relative bias) are shown in Tables \ref{tab:rel_bias_inf} and \ref{tab:abs_bias_inf}. These distributions ideally should be close to a Dirac.

The inferences using the stacked LorEMU and the stacked BNN are well centred : the relative and absolute biases are typically of a few percent, a few times smaller than what the NDE predicts for $\tau_{SF}$ and $f_X$. However, the BNN give the results with the most dispersion. The NDE shows a smaller dispersion than the BNN, but significantly more systematic absolute bias ($\lesssim 20 \%$ versus $\sim5 \%$). Overall, the emulator leads to the most accurate and consistent results. The relative bias is smaller than with the other methods, which does not appear to be caused by a systematic underconfidence of the posteriors, as the same can be said of the absolute bias. 

Since the signals are noiseless, the only sources of bias remaining are CV and imperfections in the training of the network.

{Still, for all methods and across all parameters, the biases are on average smaller than $\sim 0.3 \sigma$. The relative and absolute biases are roughly equivalent, pointing to the fact that the $1\sigma $ of the 1D posteriors are close to the grid step of the sampling of the parameter space, and the error in the position of the maximum of likelihood are on average small compared to that grid step. Although we cannot guarantee that no architectural modification would lead to better performances, a better sampling may be required to significantly improve the result.}

\subsection{Inferences on noised signals : simulation based calibration}

\subsubsection{Principle}

While the previous section gives measures of the performance of the inference methods on clean signals under the assumption of posteriors with simple shapes, a more general and rigorous validation test exists using a large number of inferences on noised signals. Assuming a perfect processing of systematics and removal of foregrounds, we apply our networks to signals distorted only by a noise realization corresponding to 100h of SKA observations. 

In the case of clean signals, we could expect the {positions of the maximum of posterior} to be (on average) close to the "true" values, i.e; the values used to generate the corresponding signals, with the remaining error caused by an imperfect training of the networks and by our inability to properly marginalize over cosmic variance.

When inferring on noised signals, even in an ideal case where the inference methods are not affected by these issues, we expect the recovered parameter to significantly deviate from the true values. Intuitively, the $N \%$ confidence contour of the posterior should contain the true values $N \%$ of the time. Robust frameworks that expand on this intuition are given by \cite{Talts2018} and \cite{Cook2006}, with the former already used in the context of 21-cm inference by \cite{Prelogovic2023}. We follow this approach and apply the Simulation Based Calibration (SBC) of \cite{Talts2018} to our inference methods. To do so, we perform 1000 inferences on noised signals and sample the obtained posteriors $N$ times. In the case of the emulator and the NDE, the sampling is the chain of steps given by the MCMC algorithm, and in the case of the BNN, we can simply draw from the distribution parametrized by the output of the network.  We then compute for each astrophysical parameter and for each inference the rank statistics $r$ , defined here by 

\begin{equation}
    r = \sum_{n = 1}^{N} \mathbf{1}[ \theta_n^{sample} < \theta^{true} ]
\end{equation}

\noindent where $\theta_n^{sample}$ is the $n^{th}$ sample of the posterior, $\theta^{true}$ the true values of the astrophysical parameter, $\mathbf{1}$ the indicator function (which is $1$ where $ \theta_n^{sample} < \theta^{true}$ and $0$ elsewhere). In other words, the rank counts how many samples of the posterior lie below the true value of the parameter. For simplicity, we normalize this quantity and use $\Tilde{r} = \frac{r}{N}$. For each astrophysical parameter, we then compute histograms of the obtained normalized rank statistics. 

If the marginalized 1D posteriors are the true ones, then the true parameters are just another draw from the posteriors and the histograms of $r$ must be flat. {This relies on the fact that the random variable $X = \int_{\theta < \theta true}  P(\theta | D)d\theta $  (the cumulative distribution function of the 1D true posterior) will be uniformly distributed \citep[under continuity assumptions,][]{Cook2006}, and the rank statistics defined above is simply a discrete version of this.}

As a simple example to illustrate this idea, let us focus on the deciles of the posterior distributions. In the ideal case, $\theta^{true}$ will lie $10\%$ of the time in the $k^{th}$ decile, which will result in $\frac{k-1}{10} < \Tilde{r} < \frac{k}{10}$. This is naturally reflected by a flat histogram and obviously independent of the choice to use deciles instead of any other quantile. Note that in the usual case of a Gaussian distribution, this is a more general way to show that the true value falls e.g. $\approx68 \%$ of the time within $1\sigma$ of the mean of the distribution.

Conversely, if the inferred distributions are not the true posteriors, the histograms will not be flat. If they are biased, then $\theta^{true}$ is more likely to be found on one side of the distribution than the other, which means $\Tilde{r}$ will be systematically closer to $0$ or $1$ than in the ideal case, and the histogram will be tilted. If they are unbiased but over-confident, then there is an excess probability that $\theta^{true}$ will fall on the fringe of the distribution and the histogram will be $\cup$-shaped. Similarly, if they are under-confident, then $\theta^{true}$ will tend to fall closer to the centre of the distribution than what the ideal case entails, and the histogram will be  $\cap$-shaped. Therefore, the $\Tilde{r}$ histograms give us a clear way to measure the performance and caveats of our methods.

\subsubsection{SBC results and discussion}

We show the results for the three inference methods on Figs. \ref{fig:sbc_emu}, \ref{fig:sbc_nde}, and \ref{fig:sbc_bnn}. The $\Tilde{r}$ histograms are shown in blue. To better understand these results, we also generate a "mock truths" normal distribution of mean 0 and variance 1 $N(0,1)$, and compute "mock rank statistics" by sampling comparatively biased and over or under-confident normal distributions and computing the corresponding $\Tilde{r}$. {We tune the mean and variance of the biased or over/underconfident mocks to obtain histograms that match those of our real case and plot these mock histograms in orange.} As before, we exclude the very poorly recovered $f_{esc}$ from our analysis. 

\begin{figure*}[h]
    \centering
    \includegraphics[scale = 0.75]{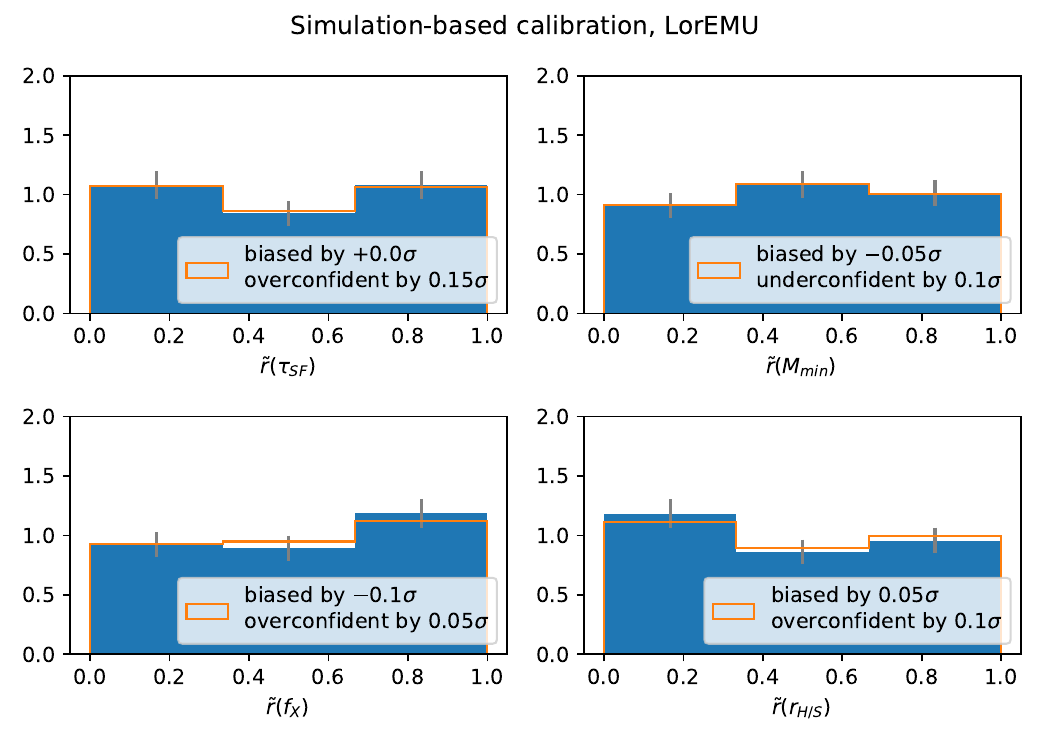}
    \caption{ $\Tilde{r}$ distributions for 1000 inferences done with the stacked emulator (\textit{blue)}. We interpret the non-flatness of the histogram by identifying it with the histogram we would obtain with a Gaussian toy model (\textit{orange)}. {The error bars indicate the standard deviation of each bin.}  }
    \label{fig:sbc_emu}
\end{figure*}

In the case of the emulator, the $\Tilde{r}$ histograms feature small systematic biases, as well as slight under- or overconfidence. These effects are of the order of $0.1 \sigma$ and consistent with sampling noise. For example, the bottom-left histogram of Fig. \ref{fig:sbc_emu} are what one would obtain if the true posterior were a $N(0,1)$ distribution, but the inferred posterior a $N(-0.1,0.95^2)$ distribution. The biases observed here are consistent with Table \ref{tab:rel_bias_inf}, with the difference explainable by a possible mismatch between the median and the mean of the posteriors i.e. their non-Gaussianity. 


\begin{figure*}[h!]
    \centering
    \includegraphics[scale = 0.75]{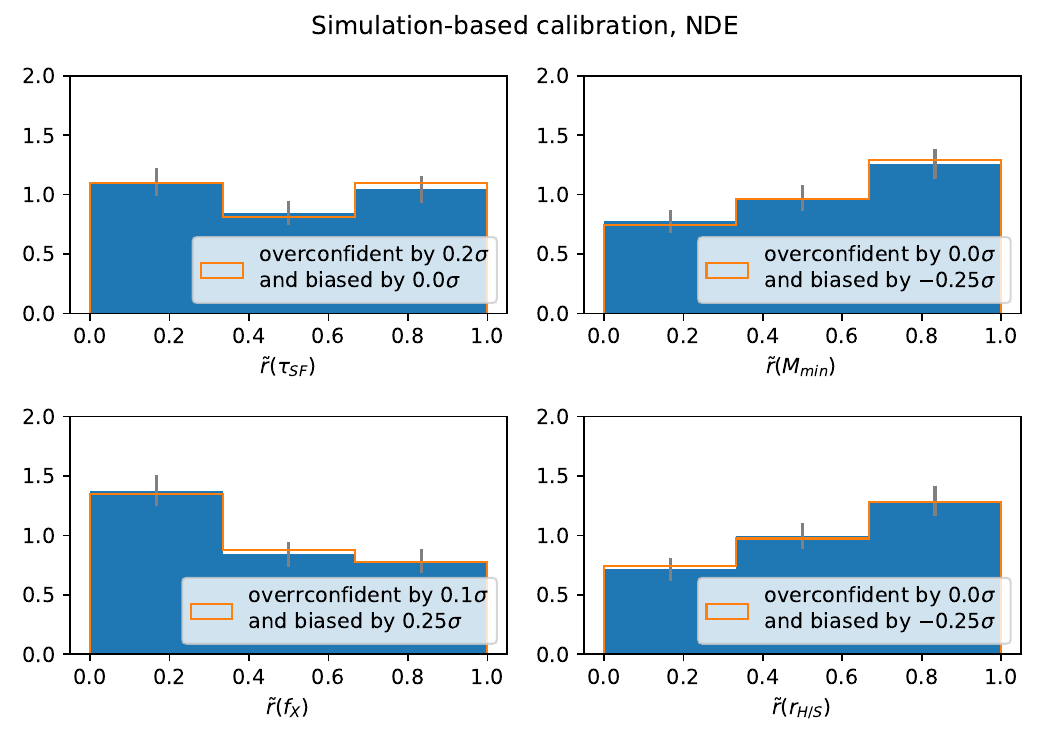}
    \caption{Same as \ref{fig:sbc_emu}  with the stacked NDE. }
    \label{fig:sbc_nde}
    \centering
    \includegraphics[scale = 0.75]{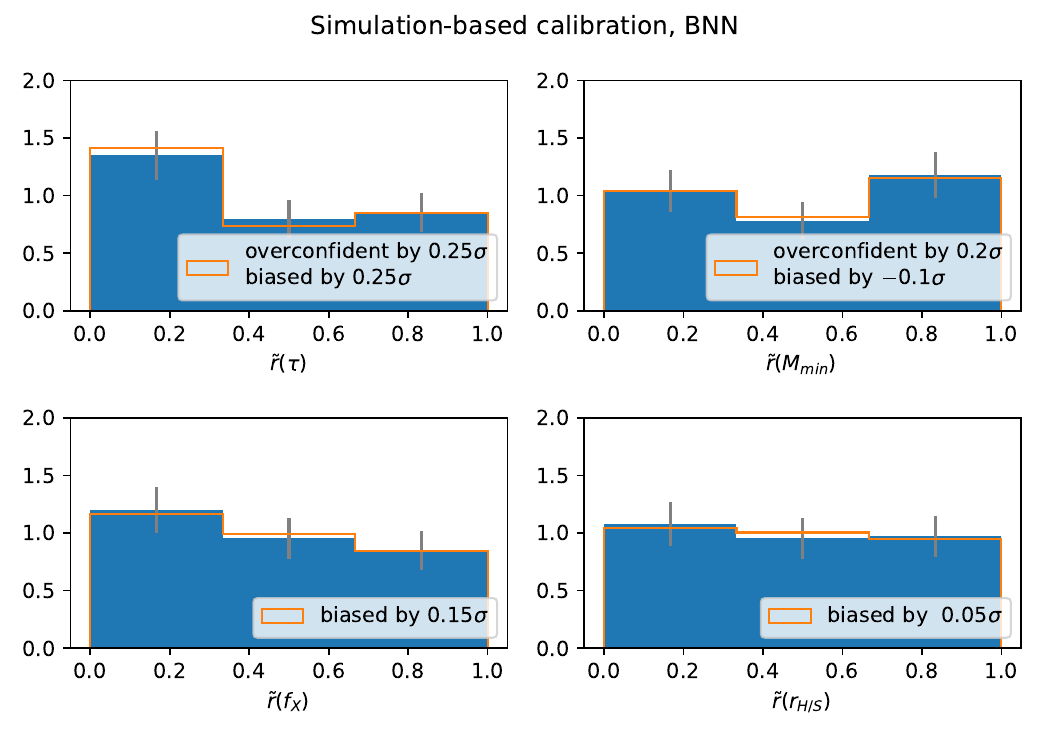}
    \caption{Same as \ref{fig:sbc_emu}  with the stacked BNN.}
    \label{fig:sbc_bnn}
\end{figure*}

\begin{table}[]
    \centering
    \begin{tabular}{c|c|c|c}
        $\delta\mu/\sigma$  & EMU & NDE & BNN  \\
   \hline
       $\tau_{SF}$  &   $-0.06 \pm 0.35$  & -$0.2 \pm 0.38$     &  $-0.05 \pm 0.62$  \\
       $M_{min}$    &  $0.04 \pm 0.32$    &  $0.04 \pm 0.35$   & $0.06 \pm 0.61 $ \\
        $f_X$       & $-0.04 \pm 0.24$    & $-0.27 \pm 0.42 $  &  $0.01 \pm 0.56 $\\
        $r_{H/S}$   & $0.0 \pm 0.27$      & $ -0.04 \pm 0.3$   & $0.13 \pm 0.62$
    \end{tabular}
    \caption{relative bias (i.e. ratio of error and standard deviation of the posterior) for all methods and parameters. }
    \label{tab:rel_bias_inf}
\end{table}

\begin{table}[]
    \centering
    \begin{tabular}{c|c|c|c}
        $\delta\mu/\Delta_{grid}$  & EMU & NDE & BNN  \\
   \hline
       $\tau_{SF}$  &   $-0.05 \pm 0.38$  & -$0.2 \pm 0.38$     &  $0.03 \pm 0.71$  \\
       $M_{min}$    &  $-0.04 \pm 0.25$    &  $0.0 \pm 0.2$   & $0.02 \pm 0.44 $ \\
        $f_X$       & $0.01 \pm 0.23$    & $-0.15 \pm 0.2 $  &  $0.0 \pm 0.33 $\\
        $r_{H/S}$   & $0.07 \pm 0.47$      & $ -0.0 \pm 0.18$   & $-0.09 \pm 0.45$
    \end{tabular}
    \caption{relative bias  (i.e. ratio of error and grid step) for all methods and parameters.}
    \label{tab:abs_bias_inf}
\end{table}


{There again, the NDE and the BNN underperform in comparison, as shown on Figs. \ref{fig:sbc_bnn} and \ref{fig:sbc_nde}. The $\Tilde{r} $ histograms display higher bias (up to $\sim 0.25\sigma$) and in a few cases, under- and overconfidence. Given the histograms of the previous section, the worse performance of the SBI methods in the SBC test are not unexpected. 
Note that the histograms in the previous sections measure the difference between the true parameters and the position of the maximum of likelihood, while what we interpret as a bias when using the SBC characterizes the median of the distribution. Furthermore, we  extract quantitative information (bias and over/underconfidence) from the SBC by comparing our histograms with what we would obtain assuming that the true and recovered posteriors are Gaussian (with different parameters). It is possible that if the recovered posterior differs from the true one by e.g. its skewness or kurtosis, we would interpret this as bias and over/underconfidence. Since the biases measured using the maxima of likelihood are different from the ones measured using SBC (hence, using medians), we expect non-Gaussian posteriors (see the non-Gaussian posterior recovered by the BNN on Fig. \ref{fig:all_inf}, and differences in the biases measured on Table \ref{tab:abs_bias_inf}, \ref{tab:rel_bias_inf} and Fig. \ref{fig:sbc_bnn}). 
Still, it appears that the SBI methods display larger  errors and LorEMU leads to consistently better results. Recall that the point of using the SBI networks is to describe the likelihood and posterior with mixture models. The added flexibility may in theory allow the to fit the true, non-Gaussian target distribution, but it also means it is harder to constrain and require more training samples during training to reach the optimal performance. 
Moreover, in a regime where thermal noise is the dominant stochastic process, the assumed Gaussian likelihood of LorEMU may be very close to the true likelihood. Finally, the architectures of the SBI networks are more complex than the emulator, and algorithmic improvements remain possible. 

Although the setups are different, \cite{Prelogovic2023} explore a 5D parameter space with 21cmFAST and train Gaussian Mixture Model NDE to fit the likelihood of their model (with a noise corresponding to 1000h of SKA integration, while we use 100h). They notice a massive improvement in performance between training on $\sim 5700$ examples and on  $\sim 28000$ examples. \textsc{Loreli} II contains "only" 10000 simulations, it is therefore possible that our training sample does not contain enough simulations to benefit from the possibility of fitting the true likelihood/posterior with neutral networks. In other words, in our setup, we obtain better results with a well-evaluated approximate likelihood rather than with a comparatively poorly constrained likelihood or posterior. Still, the errors we observe are of the order of a fraction of the impact of the thermal noise and we can be reasonably confident in the results of our three inference pipelines.}

\section{Inference on real data}\label{sec:inf_hera}

\subsection{Inference on HERA upperlimits}

\begin{figure*}
    \centering
    \includegraphics[scale = 0.66]{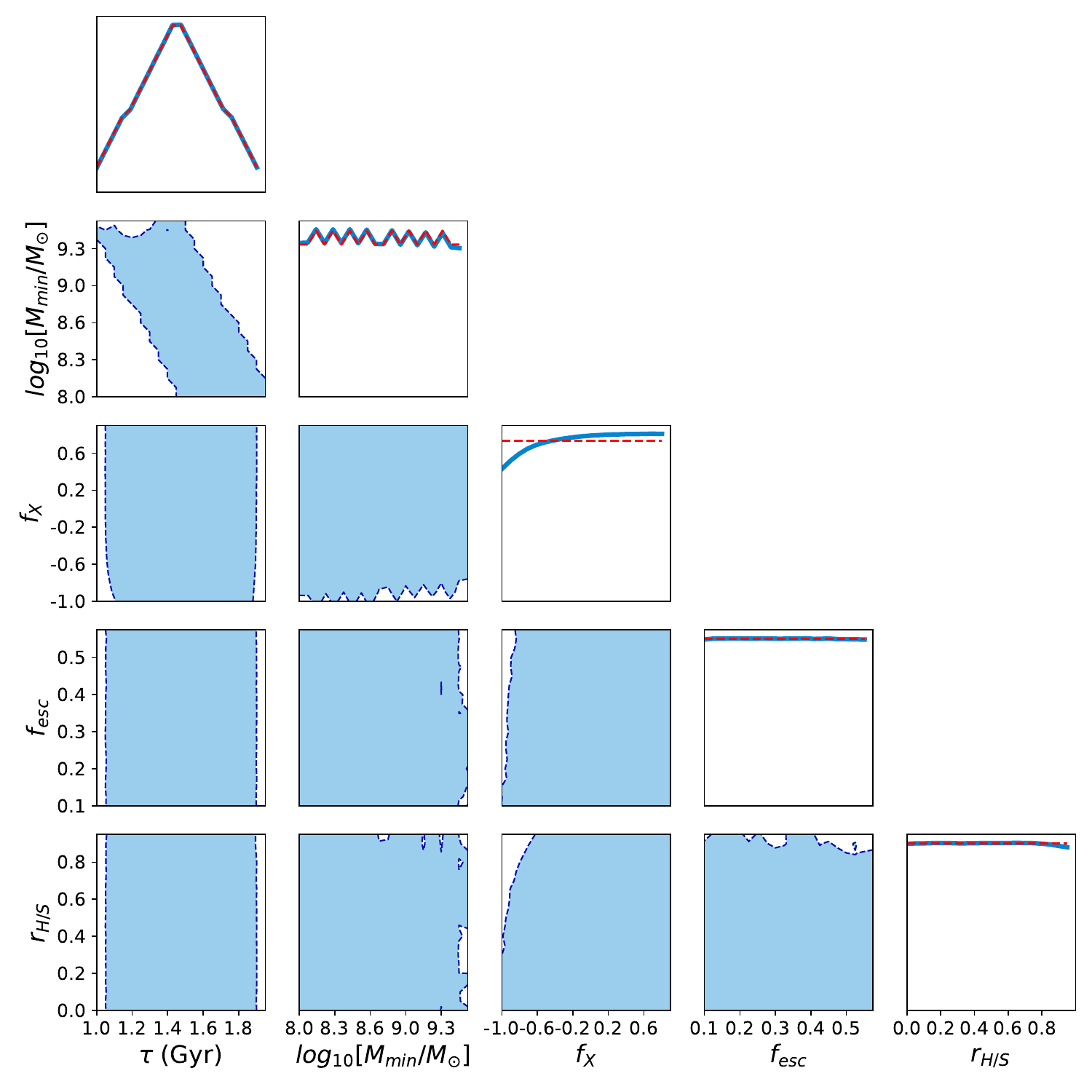}
    \caption{Inference on the latest HERA upper limits using the stacked LorEMU. Priors are shown in red, and the posteriors $95\%$ contours in blue. The posteriors over the star formation parameters are identical to the corresponding priors, but in the case of $f_X$, they show a significant departure from the prior. Models with low X-ray intensity ($log_{10} f_X \lesssim -0.6$) are deemed less likely to match the observations by HERA.}
    \label{fig:inf_hera}
\end{figure*}

Upper limits on the power spectrum produced by the HERA interferometer and given in \cite{Abdurashidova2022a} are close to the most intense signals in the \textsc{Loreli} II database and might therefore constrain our models. All the inferences shown in the previous sections were done not on upper limits but on a clean simulated signal or on a mock detection, i.e. a simulated signal to which a noise realization was added. In the case of upper limits, the target of inference is assumed to be the 21-cm signal, a noise realization, and systematics (such as foreground residuals and RFI) affecting the signal. While we model the effect of the thermal noise as a Gaussian random variable, the systematics affect the signal in an unknown, non-trivial way. To take them into account, the authors of \cite{Abdurashidova2022a, TheHERACollaboration2022} assume that they form a function $S$ that adds an unknown amount of power in each $(k,z)$ bin independently and regardless of the signal itself (and therefore of the astrophysical parameters). Hence, the likelihood in Eq.\eqref{eq:nondiag_likelihood} becomes

\begin{equation}\label{eq:likelihood_sys}
    logL(y | \theta) = -  \frac{1}{2} \left(P_{21} - S - y(\theta) \right)^T \Sigma^{-1} \left(P_{21} - S - y(\theta )\right)
\end{equation}

\noindent where $P_{21}$ is the inference target and $y$ the simulated signal. We are only interested in the posterior on the astrophysical parameters, which suggests marginalizing over the systematics $S$. If we assume a diagonal  covariance (which is a rough approximation: it is unlikely that the systematics level in adjacent $k$ bin are uncorrelated), and integrate over $S\in [0, +\infty[$ then the log-likelihood takes the form

\begin{equation}\label{eq:diag_likelihood}
    logL(y | \theta) =  \sum_{k,z}  \frac{1}{2} \left( 1 + {\rm{erf}}\left[ \frac{P_{21}(k,z) - y(\theta, k ,z )}{ \sqrt{2} \sigma_{tot}(k,z)}  \right] \right)
\end{equation}

This new form of the explicit likelihood can easily be used with the emulator in an MCMC inference pipeline.  The result of an inference done with this method are shown on Fig. \ref{fig:inf_hera}. Perhaps unsurprisingly, given the levels of the HERA upper limits, the database is not strongly constrained. The 1D posteriors for the escape fraction and the star formation parameters closely resemble the priors, but the upper limits give some constraints on the X-ray related parameters, $f_X$ and $r_{H/S}$. Strikingly, low values of $f_X$, denoting a low level of heating and a strong 21-cm signal (both on average and in its fluctuations) {are deemed less likely to describe the HERA data}. High values of $r_{H/S}$, which imply a delayed heating as harder X-rays have a larger mean free path in the neutral IGM, are also slightly constrained, as is evident in the 2D ($f_X$, $r_{H/S}$) posterior. Hence, these upper limits start to disfavour the "cold" Reionization scenarios possible in \textsc{Licorice}. Quantitatively, it must be noted that the 95$\%$ confidence contours shown on Fig. \ref{fig:inf_hera} are the regions that contain $95\%$ of the integrated probability and are therefore very much prior dependent. While the priors are physically motivated, the design of the database was done under numerical constraint: picking larger priors would cause a sparser sampling of the selected region in parameter space and could impede the training of the networks. Hence, to an extent, the exact cut-offs of the parameters range were arbitrary. Changing them would affect the exact shape of the posterior which must therefore not be over-interpreted. However, our qualitative result that low-heating models are less likely to describe the early Universe is consistent with the conclusions of \cite{TheHERACollaboration2022}.

Note that this prior-dependency is less of an issue for the results of the mock inferences shown on e.g. Fig. \ref{fig:all_inf} as the astrophysical parameters of the inference targets are near the centre of the parameter space explored in \textsc{Loreli} II and the posteriors smoothly vanish as they approach the edges of the flat prior.  Should the signal be detected, the posteriors inferred on such detection could be dangerously close to the edges of the prior. In this case, it would be wise to increase the priors by extending the database.

{Nonetheless, Fig. \ref{fig:inf_hera} and the equivalent figure obtained using \textsc{Loreli} I in \cite{Meriot2024} show the first inferences on actual instrumental data that rely on full 3D hydro-radiative simulations.}

\section{Conclusions}

In this work, we have presented the Loreli II database, a suite of 10 000 moderate-resolution simulations of the EoR that were produced  using the latest version of the Licorice hydro-radiative simulation code.  This dataset spans a 5D astrophysical parameter space, varying star formation parameters, X-ray emissivity and spectra, and UV escape fraction. We then use the Loreli II power spectra\footnote{Available at https://21ssd.obspm.fr/} as a training set to explore and compare three machine-learning-powered inference methods. The first, already introduced with Loreli I in \cite{Meriot2024}, consists in training a network to emulate the 21-cm power spectrum of Licorice simulations and using the trained emulator in an MCMC pipeline in order to perform the inference. By averaging the prediction of 10 independently trained emulators, we reach $\sim 3 \%$ accurate emulated power spectra and obtain reasonable inference results. However, this method relies on an explicit functional form for the likelihood, which we chose to be Gaussian. As discussed in \cite{Prelogovic2023}, this Gaussian shape is an approximation. We study two other methods that rely on Simulation-Based Inference in order to relax this approximation and assess whether it allows meaningful improvements. The SBI methods we study are the use of a Neural Density Estimator to fit the implicit likelihood of the Licorice simulation code and the use of a Bayesian Neural Network to directly predict the posterior over the parameters. These methods use parametric functions to fit the desired distributions, and therefore do not assume a fixed approximate shape for the likelihood.

We insist on the fact that the performance of inference methods cannot be measured and compared by studying the inference on a single signal and must be assessed statistically. To this end, we first compare the inference results on 100 random unnoised signals. Since the instrument noise is the dominant source of stochasticity in our setup, we expect the distribution of the difference between the true values of the astrophysical parameters and the maxima of likelihood to be small. According to this simple metric, the NDE shows the most systematic bias while the BNN shows dispersed results, and the emulator performs the best. In order to provide a more robust analysis, we perform Simulation Based Calibration. We count how many times the true value of the parameters lie in each quantile of 1D posteriors across 1000 inference on noised signals. Following \cite{Talts2018}, the results should ideally be uniformly distributed. However, for all methods and parameters we observe significant, albeit weak, deviations from uniformity. To make sense of these deviations, we compare them with the deviations we would obtain assuming that the true posteriors are Gaussian and that the inferred posteriors are comparatively biased or over/underconfident Gaussian. There again, we find that the emulator performs systematically better than the other methods, with errors up to $\sim0.15\sigma$ while the BNN and NDE exhibits biases and overconfidence up to $\sim 0.25 \sigma$. Still, these systematic errors are of a fraction of a parameter grid interval, which may indicate that we are limited by the density of the training set in parameter space. As in \cite{Meriot2024}, we apply the emulator to the latest HERA upper limits and obtain weak constraints on $f_{X}$, consistently with \cite{Meriot2024} and \cite{Abdurashidova2022a}. 

In conclusion, we are able to perform inference with our 3D radiative transfer simulation codes, but whether there is in practice a benefit to using SBI instead of the emulator and the approximate likelihood to infer on 100h SKA data, even with better sampling, remains an open question. Another aspect, is that SBI could show more benefits compared to an emulator if the inference was performed on summary statistics encoding non-Gaussian information, such as the Pixel Distribution Function, some compressed version of the bispectrum, or various flavor of wavelet statistics. This will be the subject of future works.

In any case, it is likely that inference on 1000h SKA data would require a more refined parameter grid : the corresponding extent of the $68\%$ contour would be significantly smaller than the current grid step, hindering the training of the networks. Hence, \textsc{Loreli} II might suffice for us to interpret upper limits on the power spectrum, perhaps even a first detection of the signal, but long-term applications will require an upgrade to the database. This is not without hassle. For \textsc{Loreli} II, we stored $\sim 50$ snapshots for each simulation, as well as particle light cones, amounting to more than one PB of data. An expanded version of Loreli will require a severe pruning of the stored data. 

Although improving the training sample can only be beneficial, another question remains. We have measured differences between the results of inference methods applied to the same physical model, but we do not know how they compare to the differences between the results of the same inference methods applied to different physical models.

Different simulation codes are parametrized differently and posteriors on the astrophysical parameters are therefore difficult to compare, but we can infer on intermediate quantities present in all codes (the ionization fraction, the bubble size distribution, properties of the galaxy population...). \cite{Zhou2022} provide an example of a prediction of the optical depth $\tau_T$ by neural networks trained on a semi-analytical model but applied to a different model. They obtain arguably poor results, demonstrating that their neural networks imperfectly generalize to data generated using other models. This might not surprise the data scientist but is an obstacle for the physicist and the 21-cm community, as it suggests that inference on real data of model-agnostics quantities will give model-dependent results. As high-redshift observations become more common, the non-21-cm related constraints on our models increase, and these models may converge to e.g. similar source models. Still, it may well be that the impact of this "model uncertainty" exceeds the difference between inference methods that we presented in this work, and this topic may require a lot of attention from the community in the future. 
Additionally, the physical modelling in the current version of \textsc{Licorice} is not without limitations. Possible improvements include the stochasticity in the star formation model (e.g. Poisson fluctuations in the CMF), the sharpness of the ionization fronts, the modelling of subgrid recombination or including the effect of local gas velocities in the computation of the Wouthuysen-Field coupling (absent from the semi-analytical code \textsc{Spinter}).
This work is a proof-of-concept, and interpretations of a future measurement of the 21-cm signal may rely on other 3D hydro-radiative simulation codes. 

\begin{acknowledgements}
This project was provided with computer and storage resources by GENCI at
TGCC thanks to the grants 2022-A0130413759 and 2023-A0150413759 on the supercomputer Joliot Curie's ROME partition. This work was granted access to the HPC resources of MesoPSL financed
by the Region Ile de France and the project Equip@Meso (reference
ANR-10-EQPX-29-01) of the programme Investissements d’Avenir supervised
by the Agence Nationale pour la Recherche. 
\end{acknowledgements}

%
%

  \bibliographystyle{aa} 
  \bibliography{My_Collection} 

\begin{thebibliography}{51}
\expandafter\ifx\csname natexlab\endcsname\relax\def\natexlab#1{#1}\fi

\bibitem[{Abdurashidova {et~al.}(2023)Abdurashidova, Adams, Aguirre, Alexander, Ali, Baartman, Balfour, Barkana, Beardsley, Bernardi, Billings, Bowman, Bradley, Breitman, Bull, Burba, Carey, Carilli, Cheng, Choudhuri, DeBoer, {de Lera Acedo}, Dexter, Dillon, Ely, Ewall-Wice, Fagnoni, Fialkov, Fritz, Furlanetto, Gale-Sides, Garsden, Glendenning, Gorce, Gorthi, Greig, Grobbelaar, Halday, Hazelton, Heimersheim, Hewitt, Hickish, Jacobs, Julius, Kern, Kerrigan, Kittiwisit, Kohn, Kolopanis, Lanman, {La Plante}, Lewis, Liu, Loots, Ma, MacMahon, Malan, Malgas, Malgas, Maree, Marero, Martinot, McBride, Mesinger, Mirocha, Molewa, Morales, Mosiane, Mu{\~{n}}oz, Murray, Nagpal, Neben, Nikolic, Nunhokee, Nuwegeld, Parsons, Pascua, Patra, Pieterse, Qin, Razavi-Ghods, Robnett, Rosie, Santos, Sims, Singh, Smith, Swarts, Tan, Thyagarajan, Wilensky, Williams, van Wyngaarden, \& Zheng}]{TheHERACollaboration2022}
Abdurashidova, T. H. C.~Z., Adams, T., Aguirre, J.~E., {et~al.} 2023, The Astrophysical Journal, 945, 124

\bibitem[{Abdurashidova {et~al.}(2022)Abdurashidova, Aguirre, Alexander, Ali, Balfour, Barkana, Beardsley, Bernardi, Billings, Bowman, Bradley, Bull, Burba, Carey, Carilli, Cheng, DeBoer, Dexter, {de Lera Acedo}, Dillon, Ely, Ewall-Wice, Fagnoni, Fialkov, Fritz, Furlanetto, Gale-Sides, Glendenning, Gorthi, Greig, Grobbelaar, Halday, Hazelton, Heimersheim, Hewitt, Hickish, Jacobs, Julius, Kern, Kerrigan, Kittiwisit, Kohn, Kolopanis, Lanman, {La Plante}, Lekalake, Lewis, Liu, Ma, MacMahon, Malan, Malgas, Maree, Martinot, Matsetela, Mesinger, Mirocha, Molewa, Morales, Mosiane, Mu{\~{n}}oz, Murray, Neben, Nikolic, Nunhokee, Parsons, Patra, Pieterse, Pober, Qin, Razavi-Ghods, Reis, Ringuette, Robnett, Rosie, Santos, Sikder, Sims, Smith, Syce, Thyagarajan, Williams, \& Zheng}]{Abdurashidova2022a}
Abdurashidova, Z., Aguirre, J.~E., Alexander, P., {et~al.} 2022, The Astrophysical Journal, 924, 51

\bibitem[{Ade {et~al.}(2016)Ade, Aghanim, Arnaud, Ashdown, Aumont, Baccigalupi, Banday, Barreiro, Bartlett, Bartolo, Battaner, Battye, Benabed, Beno{\^{i}}t, Benoit-L{\'{e}}vy, Bernard, Bersanelli, Bielewicz, Bock, Bonaldi, Bonavera, Bond, Borrill, Bouchet, Boulanger, Bucher, Burigana, Butler, Calabrese, Cardoso, Catalano, Challinor, Chamballu, Chary, Chiang, Chluba, Christensen, Church, Clements, Colombi, Colombo, Combet, Coulais, Crill, Curto, Cuttaia, Danese, Davies, Davis, {De Bernardis}, {De Rosa}, {De Zotti}, Delabrouille, D{\'{e}}sert, {Di Valentino}, Dickinson, Diego, Dolag, Dole, Donzelli, Dor{\'{e}}, Douspis, Ducout, Dunkley, Dupac, Efstathiou, Elsner, En{\ss}lin, Eriksen, Farhang, Fergusson, Finelli, Forni, Frailis, Fraisse, Franceschi, Frejsel, Galeotta, Galli, Ganga, Gauthier, Gerbino, Ghosh, Giard, Giraud-H{\'{e}}raud, Giusarma, Gjerl{\o}w, Gonz{\'{a}}lez-Nuevo, G{\'{o}}rski, Gratton, Gregorio, Gruppuso, Gudmundsson, Hamann, Hansen, Hanson, Harrison, Helou, Henrot-Versill{\'{e}},
  Hern{\'{a}}ndez-Monteagudo, Herranz, Hildebrandt, Hivon, Hobson, Holmes, Hornstrup, Hovest, Huang, Huffenberger, Hurier, Jaffe, Jaffe, Jones, Juvela, Keih{\"{a}}nen, Keskitalo, Kisner, Kneissl, Knoche, Knox, Kunz, Kurki-Suonio, Lagache, L{\"{a}}hteenm{\"{a}}ki, Lamarre, Lasenby, Lattanzi, Lawrence, Leahy, Leonardi, Lesgourgues, Levrier, Lewis, Liguori, Lilje, Linden-V{\o}rnle, L{\'{o}}pez-Caniego, Lubin, Maci{\'{a}}s-P{\'{e}}rez, Maggio, Maino, Mandolesi, Mangilli, Marchini, Maris, Martin, Martinelli, Mart{\'{i}}nez-Gonz{\'{a}}lez, Masi, Matarrese, Mcgehee, Meinhold, Melchiorri, Melin, Mendes, Mennella, Migliaccio, Millea, Mitra, Miville-Desch{\^{e}}nes, Moneti, Montier, Morgante, Mortlock, Moss, Munshi, Murphy, Naselsky, Nati, Natoli, Netterfield, N{\o}rgaard-Nielsen, Noviello, Novikov, Novikov, Oxborrow, Paci, Pagano, Pajot, Paladini, Paoletti, Partridge, Pasian, Patanchon, Pearson, Perdereau, Perotto, Perrotta, Pettorino, Piacentini, Piat, Pierpaoli, Pietrobon, Plaszczynski, Pointecouteau, Polenta, Popa,
  Pratt, Pr{\'{e}}zeau, Prunet, Puget, Rachen, Reach, Rebolo, Reinecke, Remazeilles, Renault, Renzi, Ristorcelli, Rocha, Rosset, Rossetti, Roudier, {Rouill{\'{e}} D'orfeuil}, Rowan-Robinson, Rubin{\~{o}}-Mart{\'{i}}n, Rusholme, Said, Salvatelli, Salvati, Sandri, Santos, Savelainen, Savini, Scott, Seiffert, Serra, Shellard, Spencer, Spinelli, Stolyarov, Stompor, Sudiwala, Sunyaev, Sutton, Suur-Uski, Sygnet, Tauber, Terenzi, Toffolatti, Tomasi, Tristram, Trombetti, Tucci, Tuovinen, T{\"{u}}rler, Umana, Valenziano, Valiviita, {Van Tent}, Vielva, Villa, Wade, Wandelt, Wehus, White, White, Wilkinson, Yvon, Zacchei, \& Zonca}]{Ade2016}
Ade, P.~A., Aghanim, N., Arnaud, M., {et~al.} 2016, Astronomy and Astrophysics, 594 [\eprint[arXiv]{1502.01589}]

\bibitem[{Alsing {et~al.}(2019)Alsing, Charnock, Feeney, \& Wandelt}]{Alsing2019}
Alsing, J., Charnock, T., Feeney, S., \& Wandelt, B. 2019, Monthly Notices of the Royal Astronomical Society, 488, 4440

\bibitem[{Alsing \& Wandelt(2018)}]{Alsing2018}
Alsing, J. \& Wandelt, B. 2018, Monthly Notices of the Royal Astronomical Society: Letters, 476, L60

\bibitem[{Appel(1985)}]{Appel1985}
Appel, A.~W. 1985, SIAM J. ScI. STAT. COMPUT, 6

\bibitem[{Atek {et~al.}(2018)Atek, Richard, Kneib, \& Schaerer}]{Atek2018}
Atek, H., Richard, J., Kneib, J.-P., \& Schaerer, D. 2018, MNRAS, 479, 5184

\bibitem[{Barnes \& Hut(1986)}]{Barnes1986}
Barnes, J. \& Hut, P. 1986, Nature, 324, 446

\bibitem[{Bouwens {et~al.}(2023)Bouwens, Illingworth, Oesch, Stefanon, Naidu, {Van Leeuwen}, \& Magee}]{Bouwens2023}
Bouwens, R., Illingworth, G., Oesch, P., {et~al.} 2023, Monthly Notices of the Royal Astronomical Society, 523, 1009

\bibitem[{Bouwens {et~al.}(2021)Bouwens, Oesch, Stefanon, Illingworth, Labb{\'{e}}, Reddy, Atek, Montes, Naidu, Nanayakkara, Nelson, \& Wilkins}]{Bouwens2021}
Bouwens, R.~J., Oesch, P.~A., Stefanon, M., {et~al.} 2021, {NEW DETERMINATIONS OF THE UV LUMINOSITY FUNCTIONS FROM Z 9 TO Z 2 SHOW A REMARKABLE CONSISTENCY WITH HALO GROWTH AND A CONSTANT STAR FORMATION EFFICIENCY}, Tech. rep.

\bibitem[{Bowman {et~al.}(2018)Bowman, Rogers, Monsalve, Mozdzen, \& Mahesh}]{Bowmana}
Bowman, J.~D., Rogers, A.~E., Monsalve, R.~A., Mozdzen, T.~J., \& Mahesh, N. 2018, Nature, 555, 67

\bibitem[{Cook {et~al.}(2006)Cook, Gelman, \& Rubin}]{Cook2006}
Cook, S.~R., Gelman, A., \& Rubin, D.~B. 2006, Journal of Computational and Graphical Statistics, 15, 675

\bibitem[{Field(1958)}]{Field1958}
Field, G.~B. 1958, {Excitation of the Hydrogen 21-CM Line}

\bibitem[{Finkelstein {et~al.}(2015)Finkelstein, Ryan, Papovich, Dickinson, Song, Somerville, Ferguson, Salmon, Giavalisco, Koekemoer, Ashby, Behroozi, Castellano, Dunlop, Faber, Fazio, Fontana, Grogin, Hathi, Jaacks, Kocevski, Livermore, Mclure, Merlin, Mobasher, Newman, Rafelski, Tilvi, \& Willner}]{Finkelstein2015}
Finkelstein, S.~L., Ryan, R.~E., Papovich, C., {et~al.} 2015, The Astrophysical Journal, 810, 71

\bibitem[{Furlanetto {et~al.}(2006)Furlanetto, {Peng Oh}, \& Briggs}]{Furlanetto2006a}
Furlanetto, S.~R., {Peng Oh}, S., \& Briggs, F.~H. 2006, {Cosmology at low frequencies: The 21 cm transition and the high-redshift Universe}

\bibitem[{Gal \& Ghahramani(2016)}]{Gal2015}
Gal, Y. \& Ghahramani, Z. 2016, in 33rd International Conference on Machine Learning, ICML 2016, Vol.~3, 1651--1660

\bibitem[{Ghara {et~al.}(2018)Ghara, Mellema, Giri, Choudhury, Datta, \& Majumdar}]{Ghara2018}
Ghara, R., Mellema, G., Giri, S.~K., {et~al.} 2018, Monthly Notices of the Royal Astronomical Society, 476, 1741

\bibitem[{Greig \& Mesinger(2015)}]{Greig2015}
Greig, B. \& Mesinger, A. 2015, Monthly Notices of the Royal Astronomical Society, 449, 4246

\bibitem[{Greig \& Mesinger(2018)}]{Greig2018}
Greig, B. \& Mesinger, A. 2018, Monthly Notices of the Royal Astronomical Society, 477, 3217

\bibitem[{Ishigaki {et~al.}(2018)Ishigaki, Kawamata, Ouchi, Oguri, Shimasaku, \& Ono}]{Ishigaki2018}
Ishigaki, M., Kawamata, R., Ouchi, M., {et~al.} 2018, The Astrophysical Journal, 854, 73

\bibitem[{{J. Bouwens} {et~al.}(2016){J. Bouwens}, Aravena, Decarli, Walter, da~Cunha, Labb{\'{e}}, {E. Bauer}, Bertoldi, Carilli, Chapman, Daddi, Hodge, {J. Ivison}, Karim, {Le Fevre}, Magnelli, Ota, Riechers, {R. Smail}, van~der Werf, Weiss, Cox, Elbaz, Gonzalez-Lopez, Infante, Oesch, Wagg, \& Wilkins}]{Bouwens2016}
{J. Bouwens}, R., Aravena, M., Decarli, R., {et~al.} 2016, {ALMA SPECTROSCOPIC SURVEY IN THE HUBBLE ULTRA DEEP FIELD: THE INFRARED EXCESS OF UV-SELECTED z = 2–10 GALAXIES AS A FUNCTION OF UV-CONTINUUM SLOPE AND STELLAR MASS}, Tech. Rep.~1

\bibitem[{Jennings {et~al.}(2019)Jennings, Watkinson, Abdalla, \& McEwen}]{Jennings2019}
Jennings, W.~D., Watkinson, C.~A., Abdalla, F.~B., \& McEwen, J.~D. 2019, Monthly Notices of the Royal Astronomical Society, 483, 2907

\bibitem[{Jospin {et~al.}(2022)Jospin, Laga, Boussaid, Buntine, \& Bennamoun}]{Jospin2022}
Jospin, L.~V., Laga, H., Boussaid, F., Buntine, W., \& Bennamoun, M. 2022, IEEE Computational Intelligence Magazine, 17, 29

\bibitem[{Kendall \& Gal(2017)}]{Hortua}
Kendall, A. \& Gal, Y. 2017, in Advances in Neural Information Processing Systems, Vol. 2017-Decem, 5575--5585

\bibitem[{McLeod {et~al.}(2016)McLeod, McLure, \& Dunlop}]{McLeod2016}
McLeod, D.~J., McLure, R.~J., \& Dunlop, J.~S. 2016, Monthly Notices of the Royal Astronomical Society, 459, 3812

\bibitem[{McQuinn {et~al.}(2006)McQuinn, Zahn, Zaldarriaga, Hernquist, \& Furlanetto}]{Mcquinn}
McQuinn, M., Zahn, O., Zaldarriaga, M., Hernquist, L., \& Furlanetto, S.~R. 2006, The Astrophysical Journal, 653, 815

\bibitem[{Meriot \& Semelin(2024)}]{Meriot2024}
Meriot, R. \& Semelin, B. 2024, Astronomy and Astrophysics, 683, 24

\bibitem[{Mertens {et~al.}(2020)Mertens, Mevius, Koopmans, Offringa, Mellema, Zaroubi, Brentjens, Gan, Gehlot, Pandey, Sardarabadi, Vedantham, Yatawatta, Asad, Ciardi, Chapman, Gazagnes, Ghara, Ghosh, Giri, Iliev, Jelic, Kooistra, Mondal, Schaye, \& Silva}]{Mertens2020}
Mertens, F.~G., Mevius, M., Koopmans, L.~V., {et~al.} 2020, Monthly Notices of the Royal Astronomical Society, 493, 1662

\bibitem[{Munshi {et~al.}(2024)Munshi, Mertens, Koopmans, Offringa, Semelin, Aubert, Barkana, Bracco, Brackenhoff, Cecconi, Ceccotti, Corbel, Fialkov, Gehlot, Ghara, Girard, {Grie{\ss} Meier}, H{\"{o}}fer, Hothi, M{\'{e}}riot, Mevius, Ocvirk, Shaw, Theureau, Yatawatta, Zarka, \& Zaroubi}]{Munshi2023}
Munshi, S., Mertens, F.~G., Koopmans, L.~V., {et~al.} 2024, Astronomy and Astrophysics, 681 [\eprint[arXiv]{2311.05364}]

\bibitem[{Ocvirk {et~al.}(2020)Ocvirk, Aubert, Sorce, Shapiro, Deparis, Dawoodbhoy, Lewis, Teyssier, Yepes, Gottl{\"{o}}ber, Ahn, Iliev, \& Hoffman}]{Ocvirk2020}
Ocvirk, P., Aubert, D., Sorce, J.~G., {et~al.} 2020, Monthly Notices of the Royal Astronomical Society, 496, 4087

\bibitem[{Oesch {et~al.}(2018)Oesch, Bouwens, Illingworth, Labb{\'{e}}, \& Stefanon}]{Oesch2018}
Oesch, P.~A., Bouwens, R.~J., Illingworth, G.~D., Labb{\'{e}}, I., \& Stefanon, M. 2018, The Astrophysical Journal, 855, 105

\bibitem[{Papadopoulou {et~al.}(2016)Papadopoulou, Zampoglou, Papadopoulos, \& Kompatsiaris}]{Gal2016}
Papadopoulou, O., Zampoglou, M., Papadopoulos, S., \& Kompatsiaris, I. 2016, Thesis, 43, 72

\bibitem[{Papamakarios \& Murray(2016)}]{Papamakarios}
Papamakarios, G. \& Murray, I. 2016

\bibitem[{Prelogovi{\'{c}} \& Mesinger(2023)}]{Prelogovic2023}
Prelogovi{\'{c}}, D. \& Mesinger, A. 2023, Monthly Notices of the Royal Astronomical Society, 524, 4239

\bibitem[{Prelogovi{\'{c}} \& Mesinger(2024)}]{Prelogovic2024}
Prelogovi{\'{c}}, D. \& Mesinger, A. 2024 [\eprint[arXiv]{2401.12277}]

\bibitem[{Reis {et~al.}(2021)Reis, Fialkov, \& Barkana}]{Reis2021}
Reis, I., Fialkov, A., \& Barkana, R. 2021, {The subtlety of Ly $\alpha$ photons: Changing the expected range of the 21-cm signal}, Tech. Rep.~4

\bibitem[{Rubi{\~{n}}o-Mart{\'{i}}n {et~al.}(2008)Rubi{\~{n}}o-Mart{\'{i}}n, Betancort-Rijo, \& Patiri}]{Rubino-Martin2008}
Rubi{\~{n}}o-Mart{\'{i}}n, J.~A., Betancort-Rijo, J., \& Patiri, S.~G. 2008, Monthly Notices of the Royal Astronomical Society, 386, 2181

\bibitem[{Saxena {et~al.}(2023)Saxena, Cole, Gazagnes, Meerburg, Weniger, \& Witte}]{Saxena2023}
Saxena, A., Cole, A., Gazagnes, S., {et~al.} 2023, Monthly Notices of the Royal Astronomical Society, 525, 6097

\bibitem[{Schaeffer {et~al.}(2023)Schaeffer, Giri, \& Schneider}]{Schaeffer2023}
Schaeffer, T., Giri, S.~K., \& Schneider, A. 2023, Monthly Notices of the Royal Astronomical Society, 526, 2942

\bibitem[{Semelin {et~al.}(2023)Semelin, M{\'{e}}riot, Mertens, Koopmans, Aubert, Barkana, Fialkov, Munshi, \& Ocvirk}]{Semelin2023}
Semelin, B., M{\'{e}}riot, R., Mertens, F., {et~al.} 2023, Astronomy and Astrophysics, 672, 162

\bibitem[{Sheth \& Tormen(2002)}]{Sheth}
Sheth, R.~K. \& Tormen, G. 2002, Monthly Notices of the Royal Astronomical Society, 329, 61

\bibitem[{Singh {et~al.}(2022)Singh, {Nambissan T}, Subrahmanyan, {Udaya Shankar}, Girish, Raghunathan, Somashekar, Srivani, \& {Sathyanarayana Rao}}]{Singh2021}
Singh, S., {Nambissan T}, J., Subrahmanyan, R., {et~al.} 2022, Nature Astronomy, 6, 607

\bibitem[{Sobacchi \& Mesinger(2014)}]{Sobacchi2014}
Sobacchi, E. \& Mesinger, A. 2014, Monthly Notices of the Royal Astronomical Society, 440, 1662

\bibitem[{Springel(2005)}]{Springel2005}
Springel, V. 2005, Monthly Notices of the Royal Astronomical Society, 364, 1105

\bibitem[{Talts {et~al.}(2018)Talts, Betancourt, Simpson, Vehtari, \& Gelman}]{Talts2018}
Talts, S., Betancourt, M., Simpson, D., Vehtari, A., \& Gelman, A. 2018 [\eprint[arXiv]{1804.06788}]

\bibitem[{White {et~al.}(2018)White, Zacchei, \& Zonca}]{White2018}
White, D.~M., Zacchei, A., \& Zonca, A. 2018, {Planck 2018 : Diego 60 , O. Dor{\'{e}} 62,10 , M. Douspis 53 , A. Ducout 54,52 , X. Dupac 35 , S. Dusini 61 , G. Efstathiou 65,57 * , F. Elsner 73 , T. A. En{\ss}lin 73 , H. K. Eriksen 58 , Y. Fantaye 3,18 , M. Farhang 77 , J. Fergusson 11 , R. Fernandez-Cobos 60 ,}, Tech. rep.

\bibitem[{Wouthuysen(1952)}]{Wouthuysen1952}
Wouthuysen, S.~A. 1952, Physica, 18, 75

\bibitem[{Yoshiura {et~al.}(2021)Yoshiura, Pindor, Line, Barry, Trott, Beardsley, Bowman, Byrne, Chokshi, Hazelton, Hasegawa, Howard, Greig, Jacobs, Jordan, Joseph, Kolopanis, Lynch, McKinley, Mitchell, Morales, Murray, Pober, Rahimi, Takahashi, Tingay, Wayth, Webster, Wilensky, Wyithe, Zhang, \& Zheng}]{Yoshiura2021}
Yoshiura, S., Pindor, B., Line, J.~L., {et~al.} 2021, Monthly Notices of the Royal Astronomical Society, 505, 4775

\bibitem[{Zhao {et~al.}(2022{\natexlab{a}})Zhao, Mao, Cheng, \& Wandelt}]{Zhao2022}
Zhao, X., Mao, Y., Cheng, C., \& Wandelt, B.~D. 2022{\natexlab{a}}, The Astrophysical Journal, 926, 151

\bibitem[{Zhao {et~al.}(2022{\natexlab{b}})Zhao, Mao, \& Wandelt}]{Zhao2022b}
Zhao, X., Mao, Y., \& Wandelt, B.~D. 2022{\natexlab{b}}, The Astrophysical Journal, 933, 236

\bibitem[{Zhou \& {La Plante}(2022)}]{Zhou2022}
Zhou, Y. \& {La Plante}, P. 2022, Publications of the Astronomical Society of the Pacific, 134 [\eprint[arXiv]{2112.03443}]

\end{thebibliography}






   
  



\end{document}